\documentclass{aa}  

\usepackage{float}
\usepackage{adjustbox}
\usepackage{csquotes}
\usepackage{graphicx}
\usepackage[normalem]{ulem}
\usepackage{rotating}
\usepackage{tabularx}
\usepackage{lscape}
\usepackage{caption}
\usepackage{txfonts}
\usepackage{subcaption}
\usepackage[]{hyperref}
\hypersetup{
    colorlinks=true,
    linkcolor=blue,
    filecolor=magenta,
    citecolor=blue,
    urlcolor=blue,
    pdftitle={Overleaf Example},
    pdfpagemode=FullScreen,
    }
\usepackage{color}
\definecolor{todo}{rgb}{0.89,0.0,0.13}

\newcommand\T{\rule{0pt}{2.6ex}}       
\newcommand\B{\rule[-1.2ex]{0pt}{0pt}} 

\begin{document}

   \title{A comprehensive study of $\delta$ Scuti-type pulsators\\ in eclipsing binaries: oscillating eclipsing Algols}

\author{T. B. Pawar,\inst{1,2}\fnmsep\thanks{E-mail: pawartilak7@gmail.com}
\and
A. Miszuda,\inst{3}
\and
K. G. He{\l}miniak,\inst{1}
\and
F. Marcadon,\inst{3}
\and
A. Moharana,\inst{1,4}
\and
G. Pawar,\inst{1}
\and
M. Konacki\inst{1}
}

\institute{Nicolaus Copernicus Astronomical Center, Polish Academy of Sciences, ul. Rabia\'{n}ska 8, 87-100 Toru\'{n}, Poland
\and
Villanova University, Dept.\ of Astrophysics and Planetary Sciences, 800 East Lancaster Avenue, Villanova, PA 19085, USA
\and
Nicolaus Copernicus Astronomical Center, Polish Academy of Sciences, ul. Bartycka 18, 00-716 Warszawa, Poland
\and
Astrophysics Group, Keele University, Staffordshire, ST5 5BG, U.K.
        }


  \abstract
{
Eclipsing double-lined spectroscopic binaries (SB2s) hosting $\delta$ Scuti-type pulsators offer a unique laboratory for simultaneously constraining stellar geometry and interior structure. In this study, we present a comprehensive analysis of five oscillating eclipsing Algol (oEA) binaries. By combining high-precision, short-cadence TESS photometry with multi-epoch high-resolution spectroscopy, we derive precise stellar and orbital parameters. Frequency power spectra were obtained using residuals from binary modelling. We further investigate the evolutionary history of these systems using a grid of MESA binary evolution simulations. Our analysis suggests that the systems must have undergone either case A or case B mass transfer, with the primary components repositioned in the Hertzsprung–Russell diagram and now pulsating in the $\delta$ Scuti regime, while the cooler secondaries are underluminous and inflated, filling their Roche lobes. This study contributes to the growing catalog of well-characterised oEA systems and our understanding of the effects of mass-transfer on the fate of these short-period binaries.
}


   \maketitle
%
\captionsetup[table]{font={rm,small}}
\section{Introduction}
Eclipsing binaries (EBs) that are also double-lined spectroscopic binaries (SB2s) are among the most important tools for determining absolute stellar parameters. Their light curves (LCs) provide insights into their geometric configurations, particularly stellar radii relative to the orbital separation. Stellar inclinations are best constrained in EBs \citep{2018MNRAS.474.4322M}. However, LCs alone do not provide information about stellar masses, which are crucial for understanding stellar evolution.

Determining stellar masses requires radial velocity (RV) measurements across the full orbital phase; hence, spectroscopic follow-up is essential to obtain the absolute parameters necessary for understanding stellar structure and evolution. This task becomes particularly challenging for systems with longer orbital periods, as phase coverage of the binary's orbit requires longer time baselines and is more difficult for observations planning and scheduling \citep{2010A&ARv..18...67T}.

EBs provide highly accurate and precise mass and radius measurements, making them crucial for testing stellar evolution models. However, these fundamental parameters alone do not reveal much about a star’s internal structure, energy transport mechanisms, or chemical composition.

Stellar oscillations are driven by intrinsic mechanism that produce observable changes in star's brightness. Careful investigation of these brightness variation can thus inform us about the physics of the driving mechanism. One of the classical pulsators, known as $\delta$ Scuti-type stars. Their masses typically range from 1.5-2.0 $M\mathrm{_{sun}}$. Recent studies have found that many $\delta$ Scuti-type stars are members of EBs \citep{liakos2017I, liakos2017II, 2017MNRAS.470..915K, 2022RAA....22h5003K, 2022ApJS..263...34C, 2023MNRAS.524..619K}. These systems offer an excellent opportunity to obtain precise geometrical configurations. Moreover, they allow us to probe stellar interiors through a detailed asteroseismic analysis of the pulsating component \citep{2020Galax...8...75L, 2022MNRAS.514..622M}.

One of the main hindrance in such studies was the quality of data from the ground based telescopes. In order to study the complex pulsation spectra, a long baseline photometric dataset is required. Continuous, high-precision photometric data from space missions -- first from \textit{Kepler} \citep{2010ApJ...713L..79K} with its four-year baseline and now from the short-cadence observations provided by the Transiting Exoplanet Survey Satellite \citep[TESS;][]{2015JATIS...1a4003R} -- have greatly enhanced our ability to detect and analyse candidate pulsators in EBs.

The remarkable success of the \textit{Kepler} and TESS space missions has led to the detection of over 1,000 $\delta$ Scuti-type pulsators in EBs. However, determining the absolute parameters of these systems remains challenging, and thus, only a small subset ($\sim$10\%) has been characterised with high precision so far \citep{2024arXiv241000763L}. Several catalogs \citep{2017MNRAS.465.1181L, 2017MNRAS.470..915K, 2019MNRAS.490.4040A, 2022ApJS..263...34C, 2022MNRAS.510.1413K} and studies \citep[e.g.,][]{2018MNRAS.474.4322M} have investigated the statistical properties of these systems, revealing empirical relationships between their physical, orbital, and pulsational characteristics.

A subset of these binaries falls into the short-period, semi-detached category with extreme mass ratios, indicating a history of mass transfer. This process alters stellar lifetimes, temperatures, and luminosities, shifting their positions on the Hertzsprung-Russell diagram. Depending on whether donor is on the main sequence, post-main sequence, or Asymptotic Giant Branch (AGB) phase, mass transfer is classified as Case A, Case B, or Case C, respectively \citep{Kippenhahn1967, 1971ARA&A...9..183P, Eggleton2006}. In Case A, the donor loses mass while still burning hydrogen in its core, evolving into a helium white dwarf or a hot subdwarf, depending on factors like the core mass, efficiency of mass transfer and the mass ratio between the donor and accretor \citep{2002MNRAS.336..449H, 2003MNRAS.341..669H}. In Case B, the donor has exhausted core hydrogen, and if mass loss occurs before helium ignition, it becomes either a helium white dwarf or a helium-burning star, later evolving into a carbon-oxygen white dwarf. The accretor, in both cases, gains mass, often moving toward the subgiant or giant phase earlier than in single-star evolution. Case C mass transfer occurs when the donor star fills its Roche lobe after exhausting helium in its core. This often leads to dynamically unstable mass transfer.

Systems such as those in our sample represent semi-detached Algol-type EBs with pulsating components. These pulsators have experienced significant interaction and, as a result, differ from the typical $\delta$ Scuti stars found in detached EBs \citep{2002ASPC..259...96M}. \citet{mkrtichian2004frequency} introduced the term \enquote{oEA} (oscillating eclipsing Algol) to describe these systems. These binaries serve as excellent astrophysical laboratories for studying binary evolution on shorter timescales, examining the effects of mass transfer on pulsations, and comparing the impact of accretion on stellar evolution against traditional single-star models \citep{2002ASPC..259...96M}.

One of the primary challenges in fully characterizing these systems is the availability of high-quality RV measurements, which are essential for deriving precise stellar masses. This issue has become even more pronounced with the advent of large-scale photometric surveys that provide extensive LC data, yet remain insufficient for determining absolute parameters without complementary RV data. Without these crucial measurements, our understanding of these systems remains incomplete -- like a bottle of finely aged wine, enticingly within reach, yet sealed by the unyielding cork of missing RV observations. In most cases, absolute masses cannot be determined without RVs, leaving many systems with unresolved characteristics.

The Comprehensive Research with \'{E}chelles on the Most Interesting Eclipsing Binaries project \citep[CR\'{E}ME;][]{2012MNRAS.425.1245H, 2015MNRAS.448.1945H, 2021MNRAS.508.5687H, 2013MNRAS.433.2357R} was initiated to address this gap in RV data. The survey systematically monitored a sample of more than 350 EBs, obtaining multi-epoch spectra to refine our understanding of their fundamental properties.

In this paper, we present a detailed analysis of five oEA systems identified within the CR\'{E}ME sample. Sections 2 and 3 outline the data collection and curation process, followed by an in-depth discussion of RV and LC modeling. Using the derived system parameters, we perform frequency analysis on the residuals of the binary model fit and conduct an evolutionary analysis. Finally, we discuss our results and present our conclusions.

\section{Target Selection and Observations}
To identify potential $\delta$ Scuti-type pulsators, systems were pre-selected based on their projected minimum masses, ensuring they fell within the expected mass range for these pulsators. Among these, five systems with exceptionally low mass ratios stood out as strong oEA candidates: TIC 64437380 (HD 139774), TIC 82474821 (SW Pup), TIC 126945917 (HD 202042), TIC 10756751 (GP Cet), and TIC 37817410 (TZ Eri). These systems provide valuable opportunities to study the effects of mass transfer on pulsations and binary evolution in semi-detached configurations.

\subsection{Photometry}
\label{sec:Photometry}
TESS \citep{2015JATIS...1a4003R} provides 2-min cadence LCs for all the targets in this study. These LCs were retrieved from the Mikulski Archive for Space Telescopes (MAST) using the \textsc{lightkurve} package \citep{2018ascl.soft12013L}. This short cadence photometry ensures that we detect all the high frequency oscillations, especially beneficial for pulsators like $\delta$ Scuti-type variables. LCs obtained from multiple TESS sectors provide a longer time baseline which ensures a good accuracy on the detected frequencies. Photometric 2-min cadence data are summarised in Table~\ref{tab:obs log}.

\begin{figure*}[htbp]
    \centering
    \includegraphics[width=0.95\textwidth]{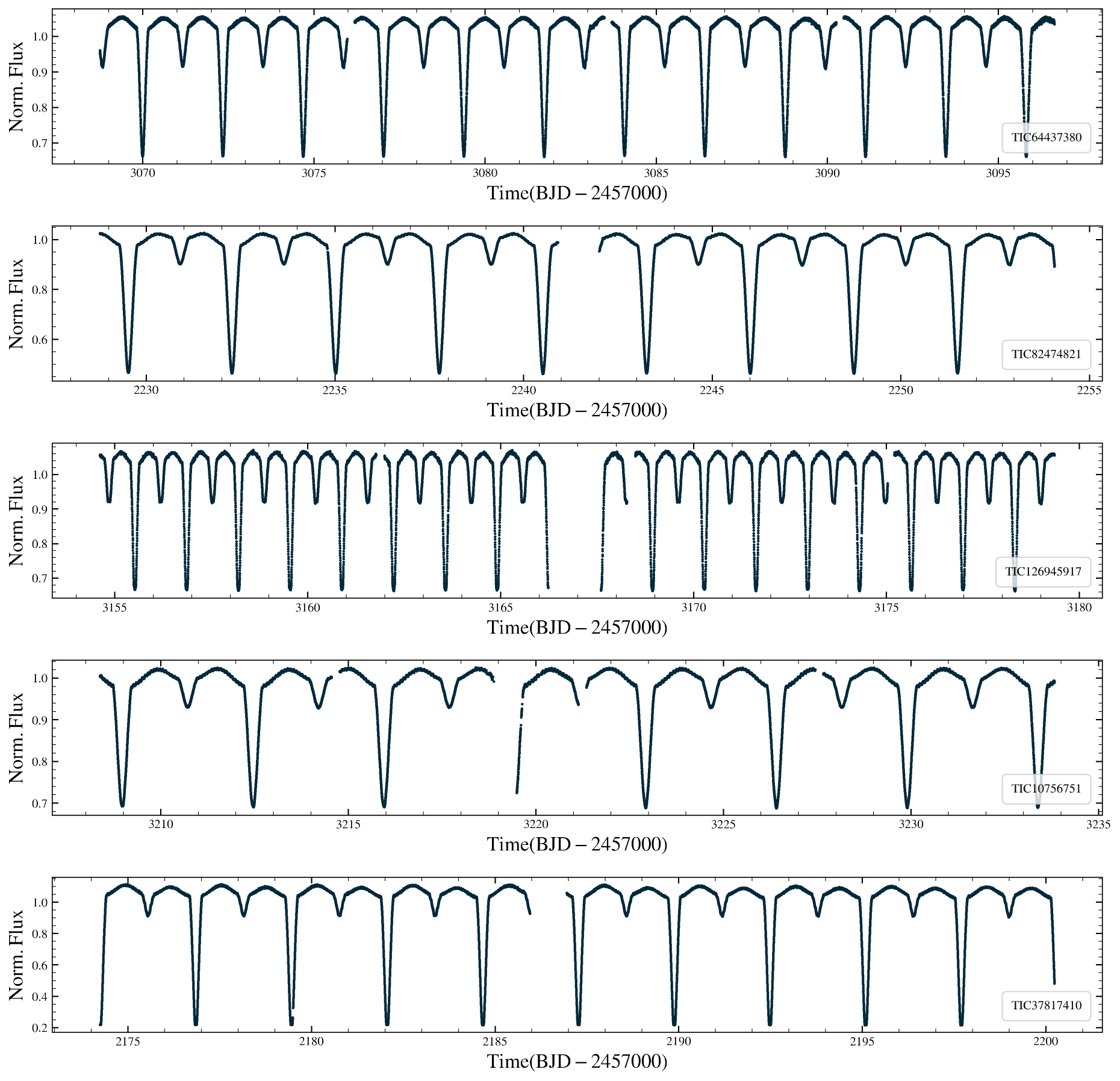}
    \caption{Short (2-min) cadence LCs of the targets, each for a single TESS sector.}
    \label{fig:LCs}
\end{figure*}

We selected the Simple Aperture Photometry (SAP) flux values for all the targets to avoid distortions in the LCs introduced through Pre-search Data Conditioning (PDC)-SAP fluxes. The data was converted to normalised fluxes and cleaned for long term trends using the \textsc{wotan} Python package \citep{2019AJ....158..143H}. LCs of all five targets used in this study are shown in the Figure \ref{fig:LCs}.

\subsection{High Resolution Spectra}
For the calculation of RVs and determination of atmospheric parameters we used high-resolution spectra, obtained with various spectrographs. Their summary is given in Table~\ref{tab:obs log}.

Most of the spectra used for analysis were taken with the FEROS spectrograph \citep{1999Msngr..95....8K}, attached to the MPG-2.2\,m telescope in La Silla. With an image slicer this instrument provides a resolution of $\sim$48000. Data were reduced with the \textsc{ceres} pipeline \citep{Brahm_ceres}, which provides wavelength-calibrated and barycenter-corrected spectra. We used 20 echelle orders, spanning 4135-6500~\AA. 

Second important source of data was the CHIRON spectrograph \citep{2013PASP..125.1336T}, attached to the 1.5\,m SMARTS telescope installed at Cerro Tololo Inter-American Observatory (CTIO). Aiming for higher efficiency, the spectrograph was used in the fiber mode, providing a resolution of $\sim$28000. Extracted and wavelength-calibrated spectra were obtained with the pipeline developed at Yale University \citep{2013PASP..125.1336T} and provided to the user. However, barycentric velocity corrections were done in-house using the {\it bcvcor} procedure within \textsc{iraf} \citep{1986SPIE..627..733T}. About forty echelle orders (4500-6500~\AA) were combined and continuum normalised (in \textsc{iraf}) to be used for RV and spectral analysis. 

One system, TIC~10756751 = GP~Cet, has been additionally observed with two other facilities: the HIDES spectrograph \citep[$R\sim55000$;][]{izu_hides}, attached to the OAO-188 telescope of the Okayama Astronomical Observatory (Japan), and the IRCS camera/spectropraph \citep[$R\sim18000$;][]{kobayashi_ircs} behind the Subaru telescope, located at Maunakea (Hawaii). Data reduction and analysis are described in details in \citet{Helminiak_hides_sb1} and \citet{2019A&A...622A.114H} for HIDES and IRCS, respectively.

\begin{table*}[htbp]
\tiny
    \centering
    \caption{Basic information about the targets.}
    \label{tab:obs log}
    \begin{tabular}{ccccccccc}
       TIC ID & Other ID & TESS sectors with & RA & Dec & $V_{\mathrm{mag}}$& GDR3\tablefootmark{a} dist. & Spectrograph(s)\tablefootmark{b} & No. of Spectra\\
        & & 2-min Photometry & (deg., ep=J2000) & (deg., ep=J2000) & & (pc) & &  \\
        \hline
        TIC~64437380 &HD~139774 & 39, 65 & 235.698351 & $-$57.510948 & 10.40 & 458.316 & 2F & 12 \\ 
        TIC~82474821 & SW~Pup & 34, 35, 61, 62 & 124.712039 & $-$42.753144 & 9.09 & 314.257 & 2F+1C & 9+6 \\ 
        TIC~126945917 & HD~202042 & 1, 27, 28, 68 &  318.732362 & $-$43.3954479 & 10.15 & 310.366 & 2F+1C & 9+5 \\ 
        TIC~10756751 & GP~Cet & 3, 42, 43, 70 & 9.229711 & $-$5.874033 & 9.90 & 481.023 & 1C+OH+SI+2F & 8+7+2+1 \\ 
        TIC~37817410 & TZ~Eri& 5, 32 & 65.418044 & $-$6.019222 & 9.61 & 299.643 & 1C & 7 \\ 
        \hline
    \end{tabular}
\tablefoottext{a}{{\it Gaia} Data Release~3 \citep[GDR3;][]{2023A&A...674A...1G}.\\}
\tablefoottext{b}{1C: SMARTS 1.5-m/CHIRON, 2F: MPG-2.2m/FEROS, OH: OAO-188/HIDES, SI: Subaru/IRCS.}
\end{table*}


\section{Spectroscopy}
RV measurements are essential for estimating the dynamical masses of stars. These RVs, derived from the Doppler shifts of spectral lines in observed spectra, directly reflect the stars' orbital motions. Additionally, estimates of effective temperature and surface gravity obtained from spectral analysis serve as valuable cross-checks for parameters such as flux ratios and relative radii derived from LC modelling. 

\subsection{Radial Velocities}
To calculate the RV values, we used the two-dimensional cross-correlation technique implemented in the \textsc{todcor} program \citep{1994ApJ...420..806Z}. Synthetic spectra used as templates were calculated using the \textsc{atlas9} model atmosphere \citep{1979ApJS...40....1K}. 
Errors were calculated using a bootstrap approach \citep{2012MNRAS.425.1245H}.

Orbital solutions for the extracted RVs were calculated using \textsc{v2fit} \citep{2010ApJ...719.1293K}. The routine estimates the orbital parameters of a double-Keplerian orbit fitted to the data, using a Levenberg-Marquardt scheme. In case of TIC~82474821, a clear Rossiter-Mclaughlin effect \citep{1924ApJ....60...15R} was observed. We choose to discard the RVs affected by this to improve the orbital solution. 

\begin{figure*}[h!]
\centering
\begin{subfigure}{0.49\textwidth}
  \centering
  \includegraphics[width=\linewidth]{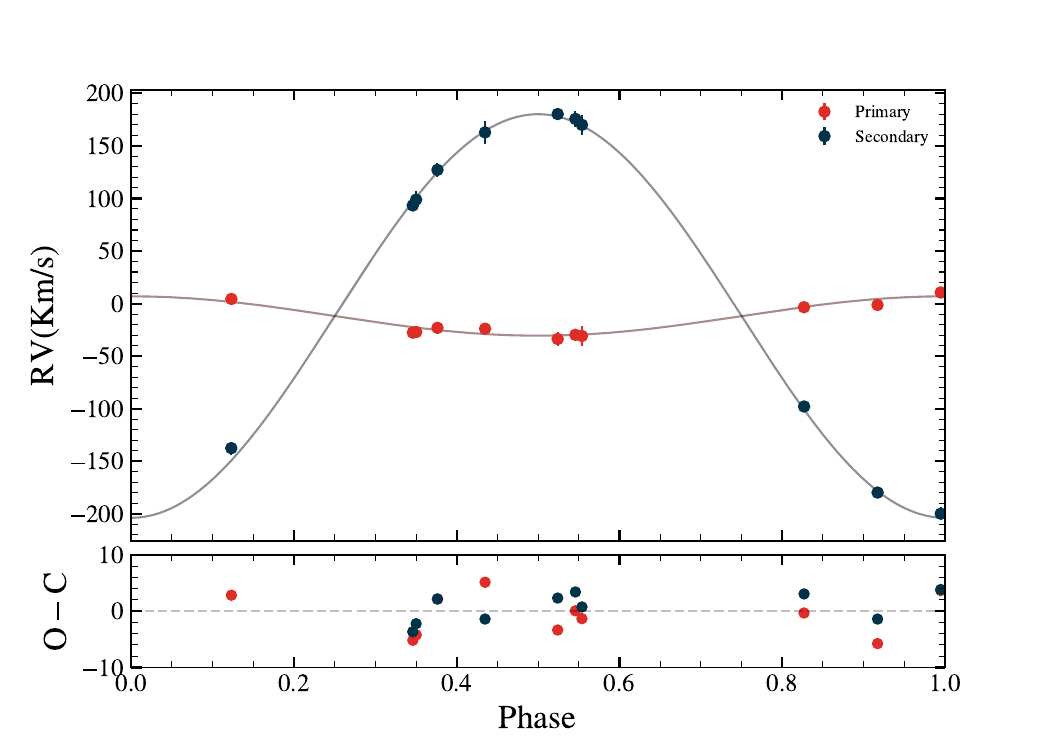}
  \caption{TIC~64437380}
\end{subfigure}
\hfill
\begin{subfigure}{0.49\textwidth}
  \centering
  \includegraphics[width=\linewidth]{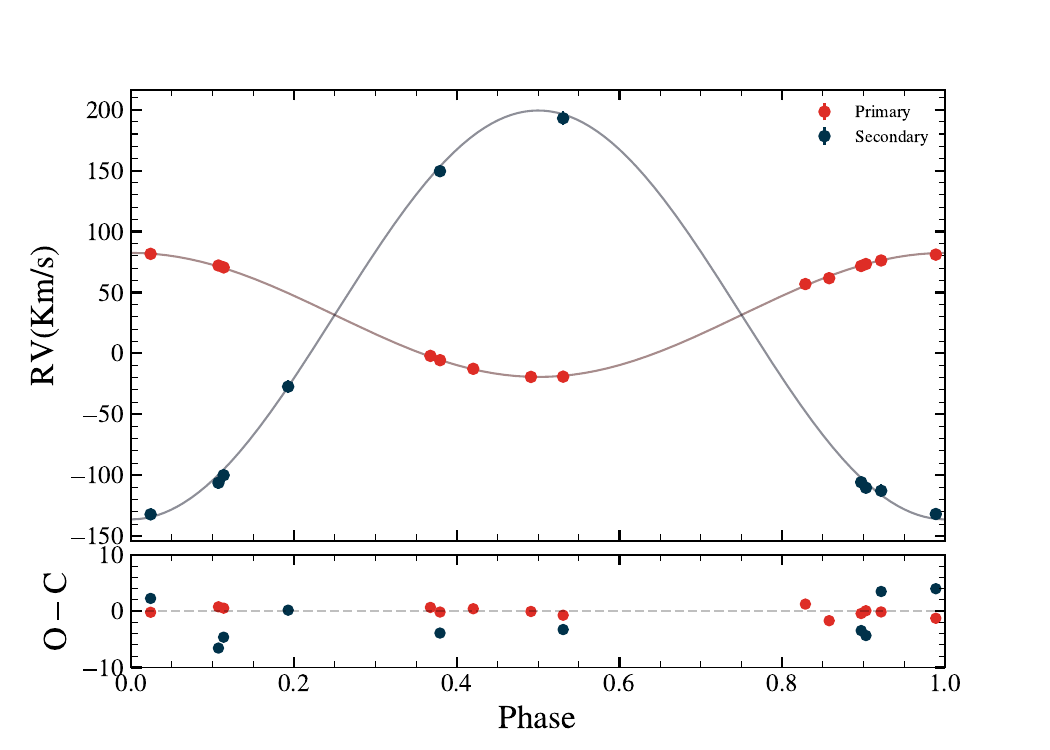}
  \caption{TIC~82474821}
\end{subfigure}
\centering
\begin{subfigure}{0.49\textwidth}
  \centering
  \includegraphics[width=\linewidth]{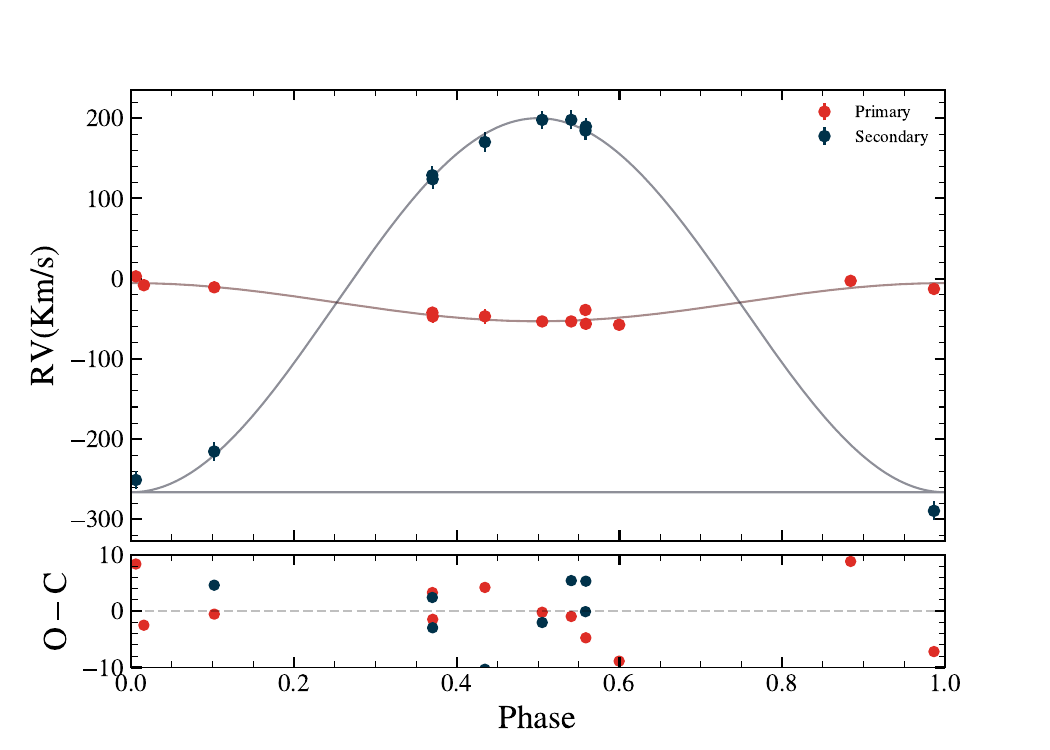}
  \caption{TIC~126945917}
\end{subfigure}
\hfill
\begin{subfigure}{0.49\textwidth}
  \centering
  \includegraphics[width=\linewidth]{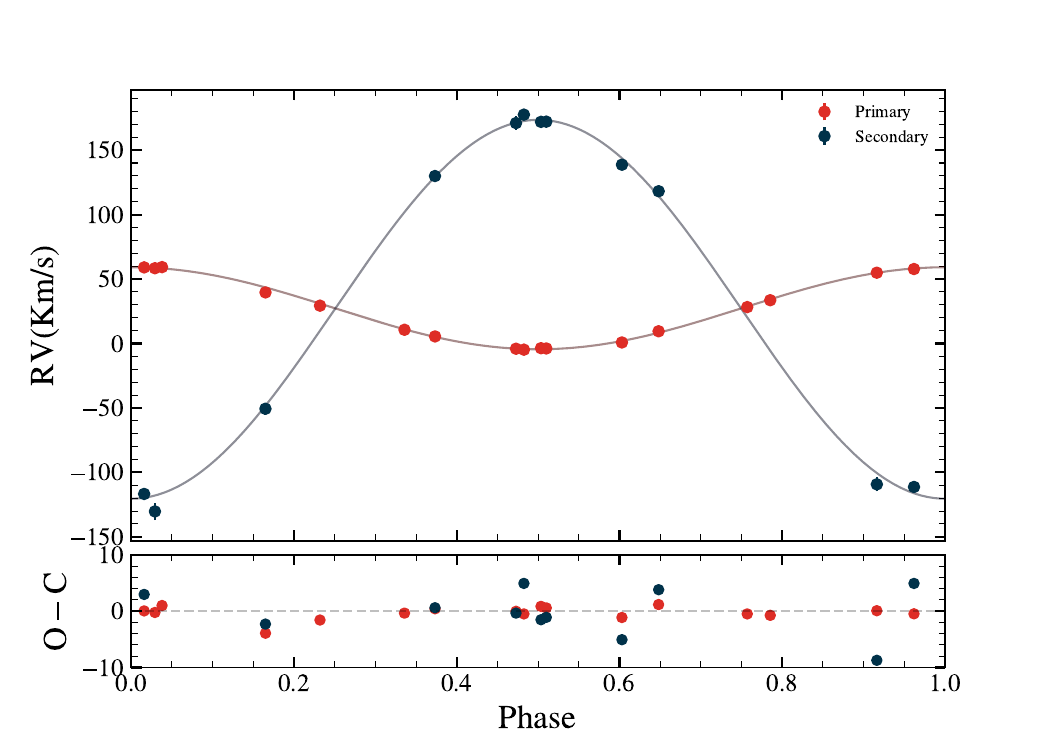}
  \caption{TIC~10756751}
\end{subfigure}
\hfill
\begin{subfigure}{0.49\textwidth}
  \centering
  \includegraphics[width=\linewidth]{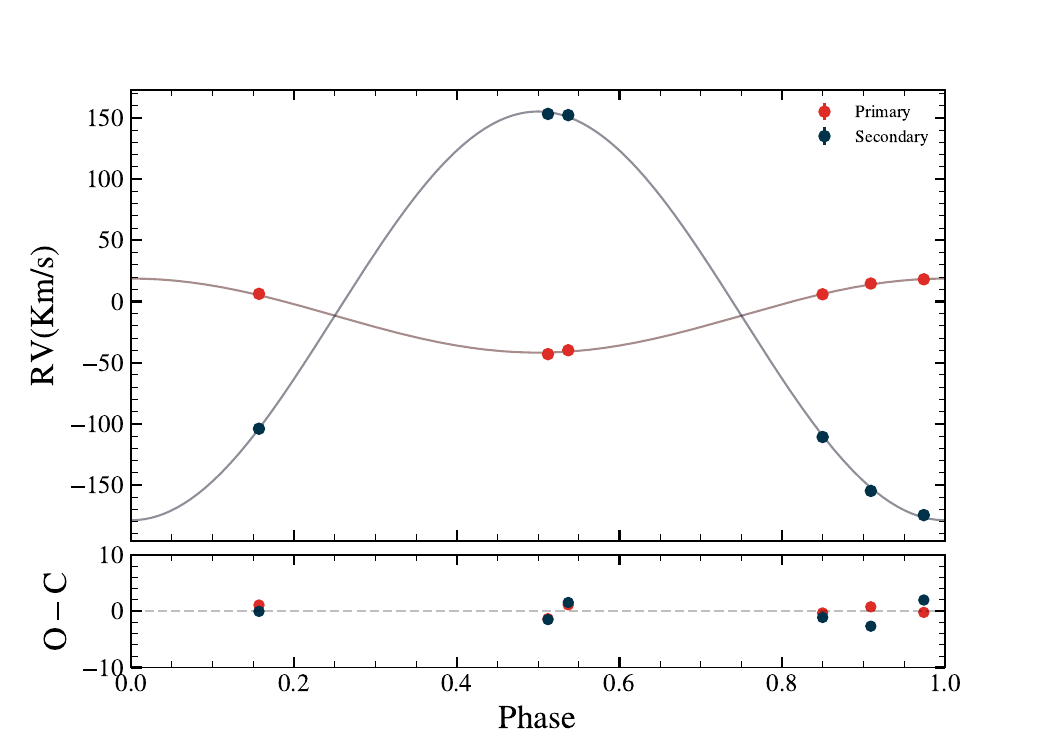}
  \caption{TIC~37817410}
\end{subfigure}

\caption{Orbital fits obtained for the RV curves for the targets. Here, primary refers to the more massive star and secondary is the less massive companion. The phase 0 is set to the time of 
the first quadrature for circular orbits.}
\label{fig:RVFits}
\end{figure*}

In general, we fit for the orbital period, $P\mathrm{_{orb}}$, time of periastron passage $T\mathrm{_p}$, systemic velocity, $\gamma$, velocity semi-amplitudes, $K\mathrm{_{1,2}}$, eccentricity, $e$ and  longitude at periastron passage, $\omega$. However, in this work we could omit the last two parameters, as the orbits were found to be circular ($e=0$) in preliminary fits. The orbital fits are displayed in Fig.~\ref{fig:RVFits}. The less massive star in all the cases is also much fainter and hence displays larger errors in the derived RV values.

\subsection{Broadening functions}
In the context of spectral disentangling and fitting, we resorted to broadening functions (BFs), mainly to obtain two necessary parameters: the rotational velocities of the stars and their light contributions. The BF depicts the spectral profiles in velocity space, containing the signatures of RV shifts across various lines, along with intrinsic stellar phenomena such as rotational broadening, spots, and pulsations \citep{1999TJPh...23..271R}. The implementation of this method is described in detail in \citet{2023MNRAS.521.1908M}.

For all five cases, the flux contribution from the fainter secondary is much smaller than from the primary. The noisy nature of the BFs for the secondary components forced us to use this analysis only for the primary stars. The rotational velocities for the primary components are provided in Table \ref{tab:Final_parameters}.

\subsection{Spectral disentangling and analysis}
The observed spectrum can be disentangled into the individual spectra of both stars, allowing for independent analysis of their atmospheric parameters. One approach is to obtain spectra at different orbital phases, where Doppler shifts cause each star’s spectrum to shift within the composite spectrum. By combining multiple spectra with their associated radial velocities (RVs), it is possible to reconstruct the individual component spectra without relying on templates.

Spectral disentangling becomes challenging when phase coverage is sparse or when one component is significantly fainter, leading to a lower signal-to-noise ratio (S/N). Unfortunately, this applies to all targets in our sample. As a result, we could not estimate the light ratio using BFs. Instead, we calculated the passband luminosities for both components in our LC model (see Sect.~\ref{sect:modeling}) and used their light ratio for the disentangling process.

We performed spectral disentangling using the \textsc{disentangling\_shift\_and\_add} code \citep{2020A&A...639L...6S, 2022A&A...665A.148S} over the 500–580 nm wavelength range. This selection avoided broad-wing features while ensuring the presence of sufficient narrow lines with adequate S/N. The used RV semi-amplitudes were obtained with \textsc{v2fit}. For further analysis, we proceeded with only the disentangled primary spectrum, continuum-corrected using the \textsc{suppnet} package \citep{2022A&A...659A.199R}\footnote{https://github.com/RozanskiT/suppnet}.


\subsection{\textsc{iSpec}}

We used the \textsc{iSpec} framework \citep{2014A&A...569A.111B,2019MNRAS.486.2075B} to determine the astrophysical parameters of the disentangled primary spectrum. If necessary, minimal continuum normalization was performed within \textsc{iSpec} using third-order splines, and slight RV shifts were also corrected.

\par
From the multiple options provided by \textsc{iSpec} for spectral fitting we use the ability to generate synthetic spectra on the go using several choices of radiative transfer codes. Specifically, we chose \textsc{spectrum} \citep{1999Msngr..95....8K} for its speed and accuracy along with \textsc{atlas9} plane-parallel model atmospheres \citep{2005MSAIS...8...14K}. We used the line lists from the Gaia-ESO Survey covering the wavelength range 420-920 nm, and solar abundances from \citet{Asplund2009}. \textsc{iSpec} uses $\chi^2$ minimization to choose the best-fit from the ones generated by \textsc{spectrum} and the observed spectra.\\

\par
During the fitting process, we fix the spectral resolution to the instrumental value, and log($g$) to the value from LC modelling (Sect. \ref{sect:modeling}) to minimise the degeneracy in fitted values. We perform the fitting procedure using the line lists best suited for determining atmospheric parameters (provided within \textsc{iSpec}) and we fit for $T_\mathrm{eff}$, $v\sin(i)$ and microturbulent velocity ($v_\mathrm{mic}$). The limb-darkening coefficients are adapted from \citet{2017A&A...600A..30C}.

\subsection{Grid Search in Stellar Paramters (GSSP)}
In order to verify the secondary temperatures obtained using the temperature ratio parameter values from the LC fitting results (Sect. \ref{sect:modeling}), we use the \textsc{gssp\_composite} module of the Grid Search in Stellar Paramters (\textsc{gssp}) software package \citep{2015A&A...581A.129T}. It uses the method of atmosphere models and spectrum synthesis, which performs a comparison of the observations with theoretical spectra from the grid. These synthetic spectra are calculated using the \textsc{SYNTHV} LTE-based radiative transfer code \citep{1996ASPC..108..198T} and a grid of atmospheric models pre-computed using \textsc{LLmodels} \citep{2004A&A...428..993S}. 

\begin{figure*}[htbp]
  \centering
  \begin{subfigure}[b]{0.95\linewidth}
    \includegraphics[width=\linewidth]{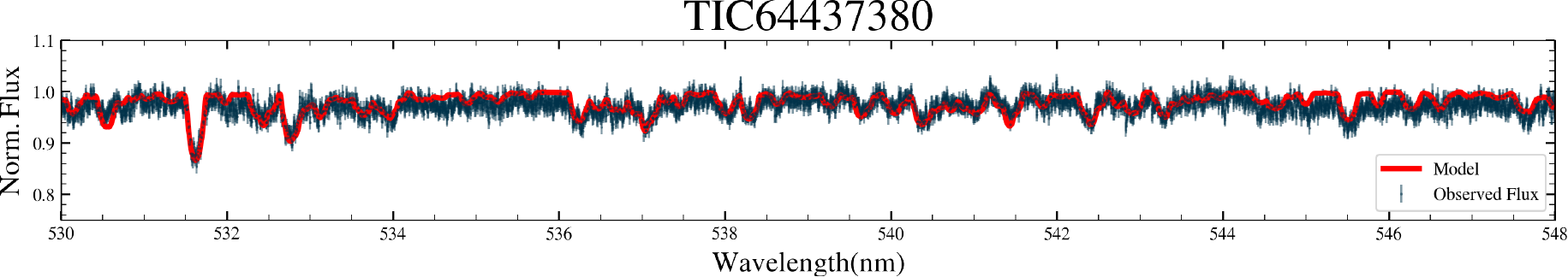}
    \label{fig:sub1}
  \end{subfigure}
  \begin{subfigure}[b]{0.95\linewidth}
    \includegraphics[width=\linewidth]{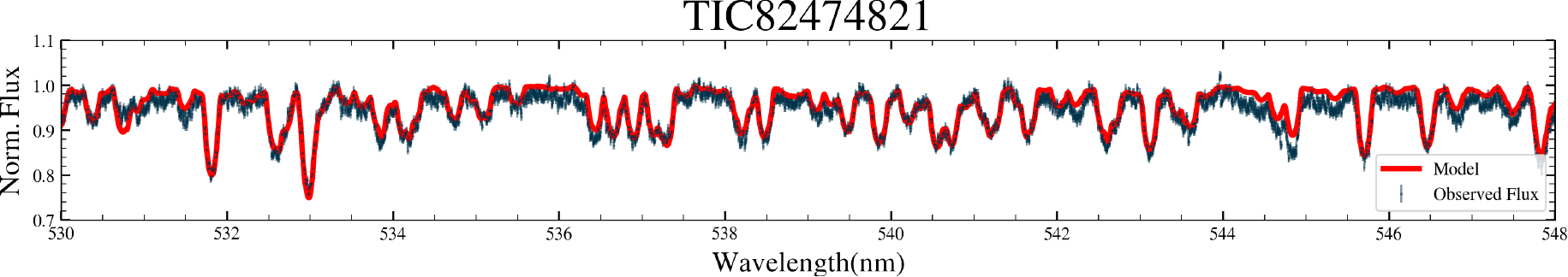}

    \label{fig:sub2}
  \end{subfigure}
    \begin{subfigure}[b]{0.95\linewidth}
    \includegraphics[width=\linewidth]{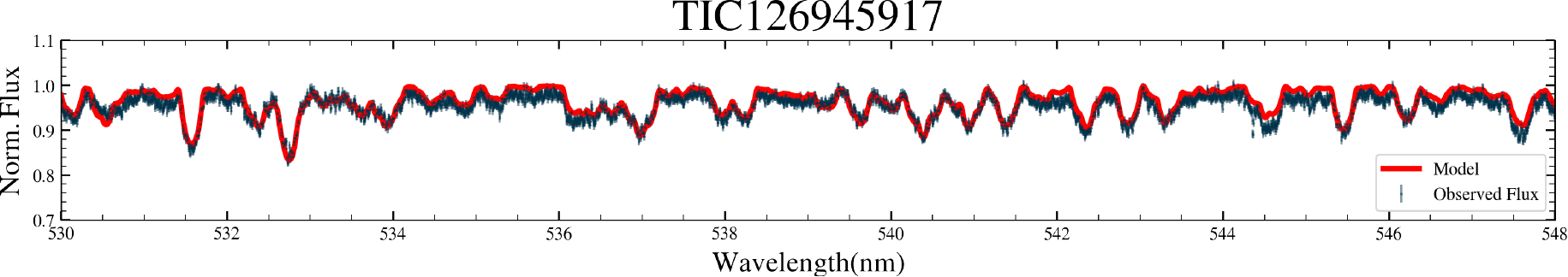}

    \label{fig:sub3}
  \end{subfigure}
  \begin{subfigure}[b]{0.95\linewidth}
    \includegraphics[width=\linewidth]{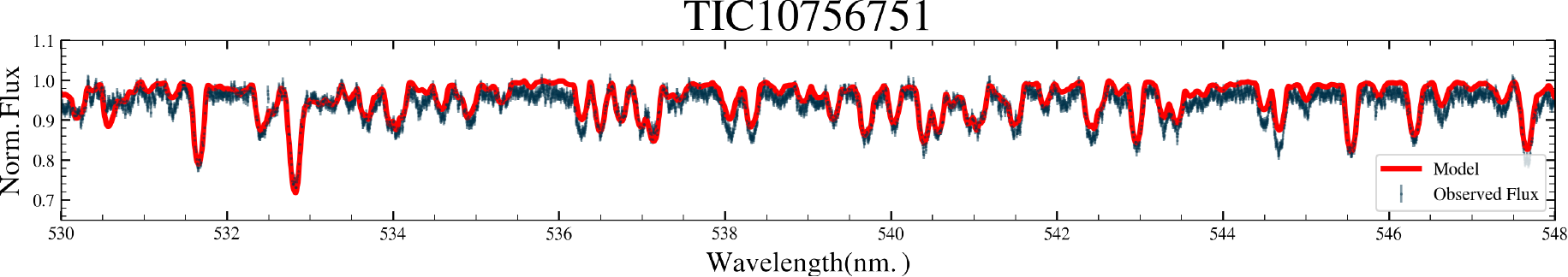}
    
    \label{fig:sub4}
  \end{subfigure}
    \begin{subfigure}[b]{0.95\linewidth}
    \includegraphics[width=\linewidth]{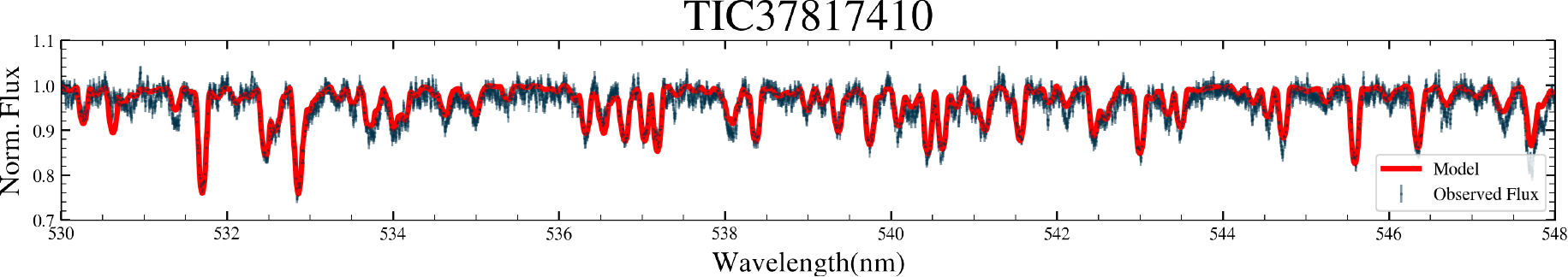}
    
    \label{fig:sub5}
  \end{subfigure}
  
  \caption{Observed composite spectrum (blue) of the superimposed targets with the \textsc{GSSP} model spectrum (in red) corresponding to the atmospheric parameters from Table \ref{tab:Final_parameters}.}
  \label{fig:ispec_results}
\end{figure*}

We initiate the primary temperature at the value estimated using \textsc{iSpec} and fit for the $T\mathrm{_{effs}}$, $v \sin (i)$, and ($v\mathrm{_{mic}}$). The best-fit values of the parameters are mentioned in Table \ref{tab:Final_parameters}. The errors are assumed to be the step-size of the parameter grid used to generate the synthetic spectra. The spectral fits are shown in Figure~\ref{fig:ispec_results}.

\section{Light curve modelling}\label{sect:modeling}
Stars in binaries with short orbital period influence each other with significant gravitational forces. These strong tidal forces cause ellipsoidal deformations that are visible in their LCs. These effects can be modelled precisely using the Roche geometry \citep{1983ApJ...268..368E}. This treatment also allows to calculate if the stars will fill their respective Roche lobe beyond the point at which mass of the star is no longer gravitationally bound to it. This allows for reproducing accurate semi-detached and contact binary configurations and their observational signatures in LCs.

The \textsc{phoebe2} \citep{2016ApJS..227...29P, 2018ApJS..237...26H, 2020ApJS..250...34C} incorporates the Roche geometry to model EBs. It also provides treatment for limb darkening laws, third light, and stellar spots, making it physically precise and thus suitable for modelling the high-precision TESS LCs of oEAs.

However, if one or both the stars in the EB harbor rapidly evolving spots, the process of obtaining a binary model becomes more challenging. This prevents us from having a single model for all the sectors as the spot parameters may affect the stellar and orbital parameter values if not taken into account. All the systems show signs of activity in the observation window. For each target we initially select the sector where minimal spot activity in the form of the O'connell effect \citep{1906MNRAS..66..123R, 1951PRCO....2...85O, 1968AJ.....73..708M} is visible in the LCs. We use these LCs to calculate our primary model.

We set the mass ratio ($q$) and the projected semi-major axis $a\sin (i)$ to the values obtained from the RV solutions. The limb darkening coefficients are determined according to the study by \citet{2017A&A...600A..30C}, while the gravity darkening coefficients are set to 1 for the primary star with radiative envelope and 0.32 for the secondary star with convective envelope. We found that using 2500 triangles to discretize the surface of each star was a good balance between accuracy and computational efficiency.

When constructing the forward model, we apply the semi-detached constraint provided in the code, ensuring that the equivalent radius reaches its maximum value, corresponding to the star filling its Roche lobe. The forward model was then optimised using a combination of Nelder-Mead and Differential Evolution optimizers within \textsc{phoebe2}. The fitted parameters include the radii ratio, effective temperature ratio, passband luminosity, and orbital inclination.

In some cases while using semi-detached configuration, we encountered convergence issues at the edge of the Roche tails. To mitigate these errors, we switched to a detached configuration, setting the radius of the secondary star as close as possible to the value of largest possible equivalent radius constrained by the Roche configuration -- without triggering errors -- before proceeding with the optimization runs.

\begin{figure*}[h!]
\centering
\begin{subfigure}{0.475\textwidth}
  \centering
  \includegraphics[width=\linewidth]{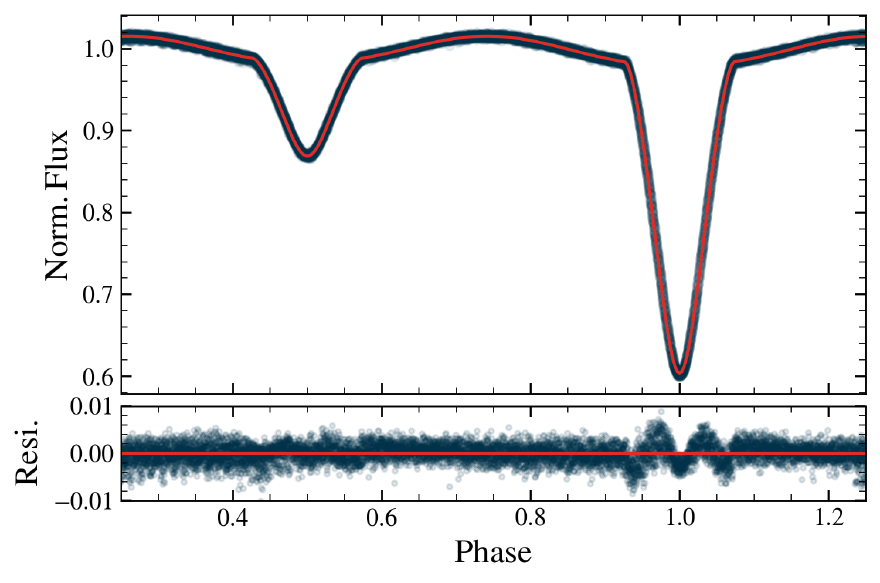}
  \caption{TIC~64437380}
\end{subfigure}
\hfill
\begin{subfigure}{0.475\textwidth}
  \centering
  \includegraphics[width=\linewidth]{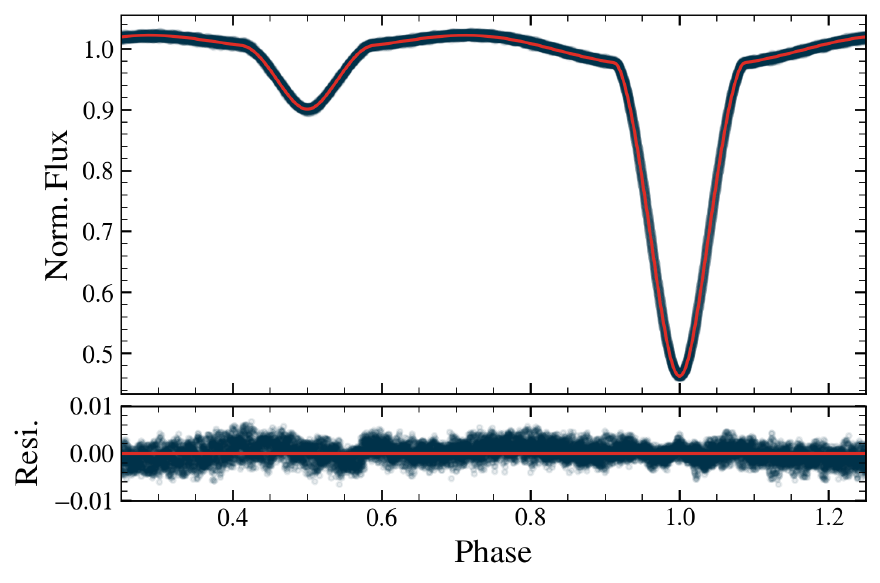}
  \caption{TIC~82474821}
\end{subfigure}
\centering
\begin{subfigure}{0.475\textwidth}
  \centering
  \includegraphics[width=\linewidth]{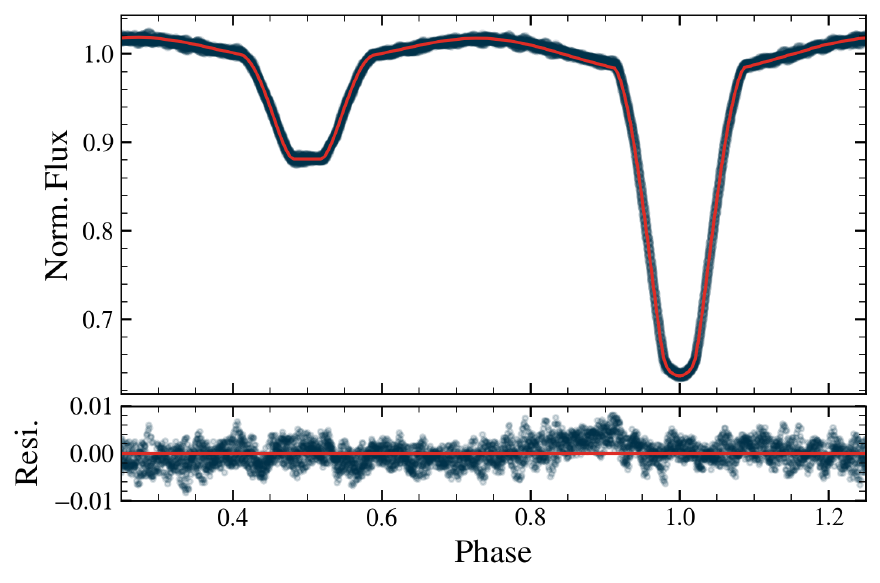}
  \caption{TIC~126945917}
\end{subfigure}
\hfill
\begin{subfigure}{0.475\textwidth}
  \centering
  \includegraphics[width=\linewidth]{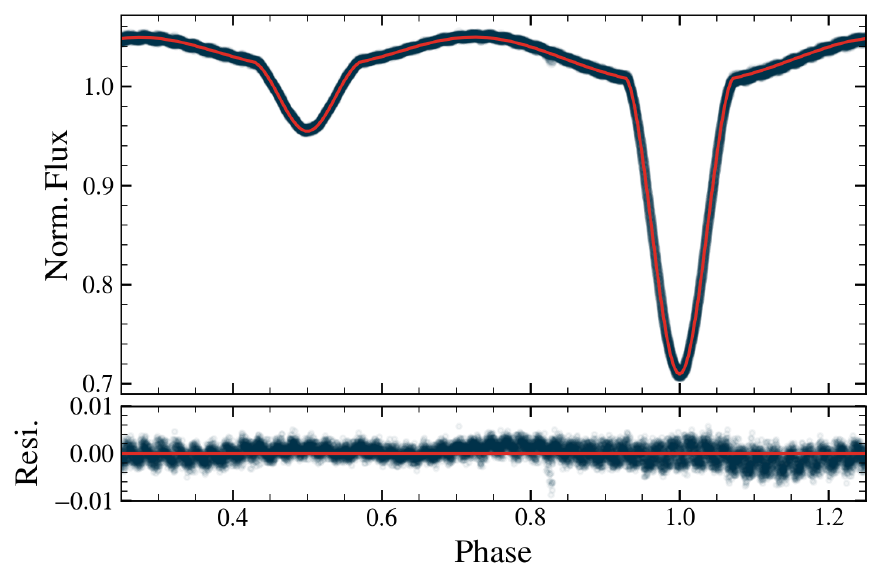}
  \caption{TIC~10756751}
\end{subfigure}
\hfill
\begin{subfigure}{0.475\textwidth}
  \centering
  \includegraphics[width=\linewidth]{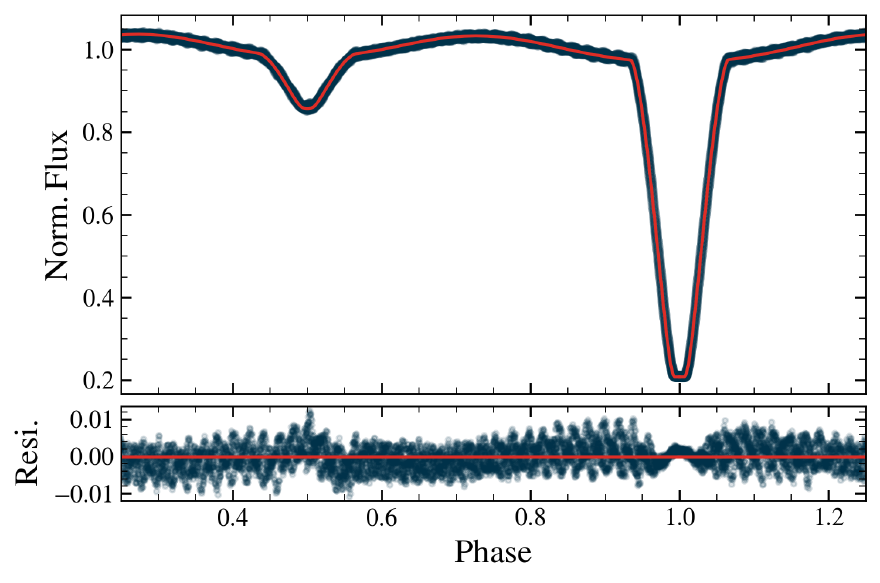}
  \caption{TIC~37817410}
\end{subfigure}

\caption{The results of LC modelling of the systems using the \textsc{phoebe2} software. Upper panel(s): Phased TESS LC (blue) with the best-fitting model superimposed (red). Lower panel(s): Residuals from the fitting procedure.}
\label{fig:LCFits}
\end{figure*}

The errors were estimated using the Markov Chain Monte Carlo (MCMC) sampling, implemented in \textsc{phoebe2} via the \textsc{emcee} sampler \citep{emcee}. The primary parameters for which we assessed uncertainties included ratio of the stellar radii, inclination, passband luminosity, and the temperature ratio. We show the results in figure \ref{fig:LCFits} and Table \ref{tab:Final_parameters}. 

\begin{table*}[h!] 
\centering
\small
    \caption{Parameters of the five oEA systems obtained using a combination of LC modelling and spectral analysis. In some cases we mention the method used to the determine the parameters in the ``()'' bracket alongside the parameter name.}
    \label{tab:Final_parameters}
    \begin{adjustbox}{max width=\textwidth}
    \begin{tabular}{lccccc}
       Parameters & TIC~64437380 & TIC~82474821 & TIC~126945917 & TIC~10756751 & TIC$~$37817410 \\
        \hline
        $T\mathrm{_0}$ (BJD-2457000)         & 2363.57069 $\pm$ 0.0002 & 2232.26734 $\pm$ 0.0005 & 3155.52250 $\pm$ 0.00001  & 3208.97018 $\pm$ 0.00002 & 1439.30929 $\pm$ 0.00045 \\
        $P_\mathrm{orb}$ [days]   & 2.34694 $\pm$ 0.00019   & 2.74738 $\pm$ 0.00001  & 1.34109 $\pm$ 0.00002 & 3.48827 $\pm$ 0.00019   & 2.60611 $\pm$ 0.00640 \\
        $i$ [$\mathrm{^o}$]           & 82.95 $\pm$ 0.15 & 79.95 $\pm$ 0.01 & 89.10 $\pm$ 0.19 & 75.40 $\pm$ 0.05  & 87.30 $\pm$ 0.01 \\
        $q$                            & 0.098$ \pm$ 0.017 & 0.303 $\pm$ 0.014 & 0.102 $\pm$ 0.017 & 0.217 $\pm$ 0.009 & 0.181 $\pm$ 0.008 \\
        $K_1$ [km\,s$^{-1}$]           & 18.7 $\pm$ 2.5 & 50.9 $\pm$ 0.5  & 23.8 $\pm$ 2.9 & 31.8 $\pm$ 0.4 & 30.3 $\pm$ 1 \\
        $K_2$ [km\,s$^{-1}$]     & 192.1 $\pm$ 4.2 & 167.8 $\pm$ 2.4 & 233.2 $\pm$ 6.2 & 146.9 $\pm$ 1.8 & 167.1 $\pm$ 1.4 \\
        $\gamma$ [km\,s$^{-1}$]  & -11.9 $\pm$ 1.6 & 31.56 $\pm$ 0.4 & -29.3 $\pm$ 2.4 & 27.3 $\pm$ 0.3 & -11.5 $\pm$ 0.7 \\
        $v \sin (i)_1$ [km\,s$^{-1}$] (\textsc{BF}) & $74 \pm 7$ & $57 \pm 1$ & $78 \pm 1$ & $52 \pm 2$ & $45 \pm 2$ \\
        $a$ [R$_\odot$]   & 9.782 $\pm$ 0.225 & 11.875 $\pm$ 0.133 & 6.815 $\pm$ 0.181 & 12.327 $\pm$ 0.125 & 10.171 $\pm$ 0.085\\
        $T\mathrm{_{eff,1}}$ [K] (\textsc{iSpec}) & 8450 $\pm$ 210 & 7450 $\pm$ 100 & 7780 $\pm$ 120 & 7300 $\pm$ 130 & 7920 $\pm$ 150 \\
        $T\mathrm{_{eff,2}}$ [K] (\textsc{GSSP}) & 5900 $\pm$ 200 & 4400 $\pm$ 200 & 5100 $\pm$ 200 & 4500 $\pm$ 200 & 4800 $\pm$ 200 \\
        $T\mathrm{_{eff,2}}/T\mathrm{_{eff,1}}$ & 0.653 $\pm$ 0.001 & 0.569 $\pm$ 0.001 & 0.709 $\pm$ 0.005 & 0.635 $\pm$ 0.005 & 0.554 $\pm$ 0.001 \\
        $M_1$ [M$_\odot$]     & 2.14 $\pm$ 0.14  & 2.39 $\pm$ 0.09 & 2.15 $\pm$ 0.17 & 1.87 $\pm$ 0.06 & 1.76 $\pm$ 0.04 \\
        $M_2$ [M$_\odot$]     & 0.208 $\pm$ 0.033 & 0.725 $\pm$ 0.019 & 0.219 $\pm$ 0.033 & 0.406 $\pm$ 0.011 & 0.319 $\pm$ 0.012 \\
        $R_1$ [R$_\odot$]     & 2.685 $\pm$ 0.010 & 3.026 $\pm$ 0.005 & 1.989 $\pm$ 0.001 & 3.244 $\pm$ 0.015 & 1.761 $\pm$ 0.001 \\
        $R_2$ [R$_\odot$]     & 1.95 (fixed) & 3.20 (fixed) & 1.24 (fixed) & 3.19 (fixed) & 2.6 (fixed) \\
        $\log(g)_1$           & 3.940 $\pm$ 0.002 & 3.822 $\pm$ 0.001 & 4.035 $\pm$ 0.004 & 3.715 $\pm$ 0.001 & 4.202 $\pm$ 0.002 \\
        $\log(g)_2$           & 3.143 $\pm$ 0.004 & 3.232 $\pm$ 0.005 & 3.458 $\pm$ 0.002 & 3.027 $\pm$ 0.001  & 3.165 $\pm$ 0.005 \\
        \hline
    \end{tabular}
\end{adjustbox}
\end{table*}

\section{Eclipse Timing Variations}
Eclipse timing variations in eclipsing binary star systems may arise from several dynamical factors. For example, large‐scale mass transfer events can alter the orbital momentum, while the gravitational influence of a tertiary companion can perturb the binary’s orbit. Additionally, stellar spots can induce timing variations; in such cases, the O–C diagram of the eclipse minima typically shows an anti‐correlation between the primary and secondary eclipses.

To search for potential signs of active mass transfer in our systems, we examined the O–C variations. For this purpose, we determined the times of the primary and secondary eclipse minima (see Appendix~\ref{appendix:A}) using the method developed by \citet{2024ApJ...976..242M}. Since historical minima timing data are unavailable for our targets, this analysis is based solely on the available TESS observations. The O–C values are computed using the following equation:
$$ \Delta\mathrm{_{obs}} = T\mathrm{_{obs}}(E) - (T\mathrm{_0} + E \cdot P), $$

\noindent where $T_0$ is the adopted initial epoch for each dataset (indicated in the tables as Cycle 0), $P$ is the orbital period, and $E$ is the cycle number.

For TIC82474821 and TIC126945917, we observed a slight anti‐correlation between the O–C values of the primary and secondary eclipse timings, suggesting that brightness asymmetries -- likely due to stellar spots -- are present. However, due to the limited data spanning only a short time period, no definitive conclusions could be drawn regarding the mass-transfer rate.
\section{Binary evolution}
The process of mass-transfer changes the most important characteristic feature of a star that determines its evolution. The state of the stars in an Algol-type system is not representative of the evolution that a similar mass star would go through. The acceptor not only gains the extra mass but also, in that process, goes through a change in the structure and composition of its outer layers. This process also changes the orbital configuration of the binary system by means of change in angular momentum.

In order to account for these effects, theoretical modelling needs to consider the binary interaction while evolving such system of stars. To perform such binary stellar evolution we used the \textsc{mesa-binary} module within the \textsc{mesa} code \citep[Modules for Experiments in Stellar Astrophysics,][version 23.05.1]{Paxton2011, Paxton2013, Paxton2015, Paxton2018, Paxton2019, Jermyn2023} to construct models under the non-rotating approximation. 

The \textsc{mesa} code builds upon the efforts of many researchers who advanced our understanding of physics and relies on a variety of input microphysics data.
The \textsc{mesa} EOS is a blend of the OPAL \citep{Rogers2002}, SCVH \citep{Saumon1995}, FreeEOS \citep{Irwin2004}, HELM \citep{Timmes2000}, and PC \citep{Potekhin2010} EOSes.
Radiative opacities are primarily from the OPAL project \citep{Iglesias1993,Iglesias1996}, with data for lower temperatures from \citet{Ferguson2005} and data for  high temperatures, dominated by Compton-scattering from \citet{Buchler1976}. Electron conduction opacities are from \citet{Cassisi2007}.
Nuclear reaction rates are from JINA REACLIB \citep{Cyburt2010} plus additional tabulated weak reaction rates from \citet{Fuller1985}, \cite{Oda1994} and \cite{Langanke2000}. Screening is included via the prescription of \citet{Chugunov2007}. Thermal neutrino loss rates are from \citet{Itoh1996}.
The \textsc{mesa-binary} module allows for the construction of a binary model and the simultaneous evolution of its components, taking into account several important interactions between them. In particular, this module incorporates angular momentum evolution due to mass transfer.
Roche lobe radii in binary systems are computed using the fit of \citet{Eggleton1983}. Mass transfer rates in Roche lobe overflowing binary systems are determined following the prescriptions of \citet{Ritter1988} and \citet{Kolb1990}.

In order to reproduce the evolution of the system, we built an extensive grid of evolutionary models. We constructed a set of varying parameters, i.e. initial orbital period, initial masses of the components, metallicity, overshooting from the convective core, mixing-length theory scaling coefficient and a fraction of mass lost during MT, with the ranges as given in Table \ref{tab:MESA_parameter_ranges}. From this set we constructed vectors of initial parameters with the assumed numerical accuracy (see last column of Table \ref{tab:MESA_parameter_ranges}) and calculated the evolutionary tracks. The parameters were chosen randomly from uniform distributions within the given ranges. All of these parameters were chosen independently of others, except for the masses.
The masses were varied between $0.5\, \rm M_{\odot}$ and $2.5\, \rm M_{\odot}$, ensuring that the mass of the primary star was larger than that of the secondary. We tested the rate of the mass transfer loss from the system as a fraction of mass that is lost from the vicinity of the accretor in a form of a fast wind during MT, in the rage from $\beta = 0.0$ to $\beta=0.3$. We used \cite{Kolb1990} mass transfer prescription and assumed a constant eccentricity $e = 0$ throughout the system’s evolution. 
Initial orbital periods in our models were chosen from 1.2 to 5 days.

\begin{table}[htbp]
   \small
    \centering
    \caption{Summary of the parameter ranges and sampling precision used in the evolutionary modeling of the five oEA systems. The table lists each parameter, its minimum and maximum values, and the precision with which it was sampled from a uniform distribution.}
    \label{tab:MESA_parameter_ranges}
    \begin{tabular}{lccc}
       Parameter & Range & Accuracy\\
        \hline
            Initial orbital period, $P_{\rm ini}\,\rm [d]$              & 1.2 - 5.0   & $10^{-5}$  \\
            Initial masses, $M_{\rm don/acc,ini}\,\rm[M_{\odot}]$       & 0.5 - 2.5   & $10^{-3}$  \\
            Metallicity, $Z$                                            & 0.01 - 0.03 & $10^{-3}$  \\
            Convective-core overshooting, $f_{\rm ov}$                  & 0.001 - 0.030  & $10^{-3}$  \\
            Mixing-length theory parameter, $\alpha_{\rm MLT}$          & 0.5 - 1.8   & $10^{-1}$  \\
            Fraction of mass lost during MT, $\beta$                    & 0.0  - 0.3   & $10^{-1}$  \\
        \hline
    \end{tabular}
\end{table}

In our evolutionary computations, we used the AGSS09 \citep{Asplund2009} initial chemical composition of the stellar matter and the OPAL opacity tables. These tables were supplemented with data from \cite{Ferguson2005} for lower temperatures. For higher temperatures, hydrogen-poor or metals-rich conditions we used C and O enhanced tables. To determine the chemical composition of stars we varied the value of metallicity $Z$, from $0.01$ to $0.03$ and used the linear scaling relations \citep[e.g.,][]{Choi2016}:
$$ Y = Y_p + \left( \frac{Y_{\rm protosolar} - Y_p}{Z_{\rm protosolar}} \right) Z, $$ with primordial helium abundance $Y_p=0.249$ \citep{PlanckCollaboration2016}, and $Y_{\rm protosolar}=0.2703$, $Z_{\rm protosolar}=0.0142$ proto-solar abundances \citep{Asplund2009}.

Convective instability was treated using the Ledoux criterion, combined with the mixing length theory based on the \cite{Henyey1965} model, employing a mixing-length parameter between $\alpha_{\rm MLT} = 0.5$ and $\alpha_{\rm MLT} = 1.8$. In regions that were stable according to the Ledoux criterion but unstable by the Schwarzschild criterion, we applied semi-convective mixing using a scaling factor of $\alpha_{\rm sc} = 0.1$, following the formalism of \cite{Langer1985}. To address regions exhibiting inversion in the mean molecular weight, such as those formed during mass accretion, we applied thermohaline mixing using the formalism of \cite{Kippenhahn1980}, with an $\alpha_{\rm th} = 1$ coefficient. As our models did not include rotational mixing, we introduced a minimum diffusive mixing coefficient of $D = 10\  \rm cm^2\,s^{-1}$ to smooth out numerical noise or discontinuities in internal profiles.
Additionally, we accounted for overshooting beyond the formal convective boundaries, using an overshooting parameter, $f_{\rm ov}$, to capture turbulent motions extending into the radiative zone. We applied an exponential overshooting scheme by \cite{Herwig2000} on the top of the H-burning core with changing value from $f_{\rm ov}=0.001$ to $f_{\rm ov}=0.03$.

We also account for the effects of stellar winds. Our models include wind mass loss, based on the prescriptions of \cite{Vink2001}, \cite{Reimers1975} and \cite{Bloecker1995}, depending on the effective temperature and the evolutionary stage of a star. For these prescriptions, we adopt efficiency coefficients of $\eta_{\rm V} = 0.1$, $\eta_{\rm R} = 0.5$, and $\eta_{\rm B} = 0.1$, respectively.

Our intention was to apply the aforementioned grid to all systems analysed in this paper. To achieve this, we established a comprehensive grid of 5 000 models, calculated based on the parameter ranges outlined in Table \ref{tab:MESA_parameter_ranges}. This grid served as a foundation for our analysis, allowing us to explore a wide spectrum of stellar configurations. From this initial grid, we carefully selected time steps for each system, that correspond to the values of the measured orbital periods within errors. To select the best fits between the observed and theoretical values we calculated the Mahalanobis distance \citep{mahalanobis1936proceedings} for each selected time step for the masses and radii. Then, for each system we selected models whose values deviated by up to 50\% from the observed equivalents, in order to extract a sample of models that enabled us to constrain the ranges of the initial parameters. This approach led to the creation of additional grids, incorporating models computed for narrower ranges with more stringent error thresholds, up to 10\%. We summarize the effects of evolutionary computations for each system in Table\,\ref{tab:MESA_models} and Figure~\ref{fig:MESA_plots}. The orange and blue lines denote the evolutionary tracks for the acceptor and donor, respectively, with dots overplotted on each track, to mark the locations of the best-fitting timestamps. The observed positions are indicated by the error boxes. The lower panels display the evolution of masses, radii and the orbital period with the x-axes representing the evolutionary time ($\tau$) normalized to the current age ($\tau_\mathrm{{sys}}$) of the corresponding best-fit model. We also provide short descriptions of the results below. 

\textbf{TIC\,82474821}:
We have identified 16 models that match the observed parameters within 10\% of its components' masses and radii. 
Some solutions suggest Case A MT, but the majority indicates that he system was formed via non-conservative case B MT, and is 0.5-3\,Gyr old. During the evolution, the initially less massive component accretes mass from it's companion, leading to mass ratio reversal and to expansion of the convective core. The donor models align at the ridge of the ending phase of mass transfer, after the exhaustion of core hydrogen, and before the onset of hydrogen burning in the envelope. Depending on the MT case, the accretor models point to either early stages of core rejuvenation (case B, central hydrogen abundance increases to approx. 0.7) or much later stages of evolution, with $\mathrm{H} \approx 0.2-0.3$ in the case of type A mass transfer.

\textbf{TIC\,10756751}:
All of the models we were able to find fit within 15\% of the measured masses and radii of the components. According to these models, TIC\,10756751 is likely to have been created from a binary with relatively small orbit, with $P_{\rm orb} \approx 1.2-1.6$\,d, via almost conservative case A mass transfer, with up to 10\% of transferred mass lost from the system. The system is relatively old, i.e. 1.86 - 8.13\,Gyr, with the accretor close to terminal age main sequence (TAMS; $\mathrm{H} \approx 0.03-0.3$), and the donor close to becoming a helium-core pre-white dwarf.

\textbf{TIC\,37817410}:
We have found 13 models that fit within 15\% of the measured masses and radii of the components, and point to $\tau_{\rm sys} \approx 1.3 - 11.2$\,Gyr. The initial conditions of TIC\,37817410 binary are unclear, as wide ranges of initial orbital periods (1.2 - 2.5\,d) and initial masses (1-1.83\,M$_{\odot}$ for the donor, 0.5-0.9\,M$_{\odot}$ for the accretor) are possible. 
However, some of the best fitting models, point to the over-solar metallicities. In these models, the donor, after transferring a substantial amount of mass to its companion, undergoes an unstable CNO burning cycle in its envelope. This process results in a series of outbursts, occasionally causing the donor to exceed its Roche lobe and triggering brief episodes of additional mass transfer to the companion.

\textbf{TIC\,64437380} and \textbf{TIC\,126945917}:
The best evolutionary models that we were able to identify for TIC\,64437380 and TIC\,126945917 fit only up to 50\% of the components' masses and radii. We believe that the difficulties in obtaining better fits arise from the parameter ranges chosen for calculating evolutionary tracks, particularly the value of the initial orbital period for the former system and secondary's initial masses for the latter system (see Table~\ref{tab:MESA_parameter_ranges}). The models we found cluster near the lower limits of the assumed orbital period and masses of the secondary, and the resulting classification of the mass transfer as type B seems to support our hypothesis. We consider that lowering the lower limit of $P_{\rm orb}$ and $M_{\rm acc,ini}$ could lead to the identification of additional models with case A mass transfer, which might improve the overall quality of the fits.

The overall analysis supports the general evolutionary type of each system presented in this paper. After mass transfer starts in, either, case A or B scenario, the initially less massive accretor gains mass, builds and expands its convective core. On the other hand, the donor after loosing a significant fraction of its mass still retains sufficiently thick hydrogen envelope to enter the red giant phase, after which the turbulent CNO burning will cause a series of outbursts, completely washing-out any hydrogen left to, eventually, become a He white dwarf.

\begin{table*}[htbp]
   \small
    \centering
    \setlength{\tabcolsep}{2pt}
    \caption{The results of the evolutionary computations summarizing the initial and final parameters of the models. The models are sorted by fit quality.}
    \label{tab:MESA_models}
    \begin{tabular}{lllllllll|lllllllc}
        \hline\hline 
        &\multicolumn{7}{c}{\textsc{Initial parameters}} & \multicolumn{7}{c}{\textsc{Final parameters}} \T\\
        & Mod.\,\# & $P_{\rm ini}$ & $M_{\rm don,ini}$ & $M_{\rm acc,ini}$ & $Z$  & $f_{\rm ov}$ & $\alpha_{\rm MLT}$ & $\beta$ & $M_{\rm acc}$ & $R_{\rm acc}$ & $\log T_{\rm eff}^{\rm acc}$ &  $M_{\rm don}$ & $R_{\rm don}$ & $\log T_{\rm eff}^{\rm don}$ & $\log \tau_{\rm sys}$ & MT Case \B\\
        & & $\rm [d]$ & $\rm[M_{\odot}]$ & $\rm[M_{\odot}]$ &  &  &  &  & $\rm[M_{\odot}]$ & $\rm[R_{\odot}]$ &  & $\rm[M_{\odot}]$ & $\rm[R_{\odot}]$ & & $\rm [yr]$ & \\
        \hline
        \multicolumn{3}{l}{TIC\,82474821\T\B} \\    
        & 1 & 1.36039 & 2.061 & 0.958 & 0.028 & 0.022 & 1.3 & 0.13 & 2.169 & 2.931 & 3.934 & 0.66 & 3.289 & 3.662 & 9.19 & A \\
        & 2 & 2.97347 & 2.099 & 0.629 & 0.030 & 0.027 & 1.5 & 0.02 & 2.041 & 2.878 & 3.99 & 0.651 & 3.29 & 3.69 & 8.96 & B \\
        & 3 & 1.33773 & 2.148 & 0.964 & 0.017 & 0.014 & 0.5 & 0.19 & 2.162 & 2.964 & 3.914 & 0.661 & 3.284 & 3.577 & 9.18 & A \\
        & 4 & 1.26583 & 1.552 & 1.140 & 0.025 & 0.023 & 0.8 & 0.03 & 1.995 & 2.947 & 3.889 & 0.66 & 3.288 & 3.588 & 9.47 & A\\
        & 5 & 2.69152 & 2.204 & 0.709 & 0.010 & 0.018 & 1.1 & 0.11 & 2.055 & 2.863 & 3.979 & 0.687 & 3.348 & 3.663 & 8.91 & B\\
        & 6 & 3.79526 & 2.465 & 0.620 & 0.028 & 0.011 & 1.0 & 0.21 & 2.041 & 3.04 & 4.001 & 0.662 & 3.311 & 3.633 & 8.7 & B\\
        & 7 & 2.39379 & 2.481 & 0.715 & 0.014 & 0.015 & 1.0 & 0.26 & 2.089 & 2.781 & 3.978 & 0.62 & 3.23 & 3.667 & 8.78 & B\\
        & 8 & 3.10770 & 2.152 & 0.657 & 0.014 & 0.010 & 0.7 & 0.06 & 2.028 & 2.839 & 3.958 & 0.689 & 3.349 & 3.596 & 8.93 & B\\
        & 9 & 1.30167 & 1.922 & 0.933 & 0.018 & 0.027 & 1.5 & 0.02 & 2.199 & 2.812 & 3.937 & 0.62 & 3.214 & 3.678 & 9.33 & A\\
        & 10 & 3.17406 & 2.312 & 0.663 & 0.013 & 0.008 & 0.9 & 0.16 & 2.032 & 3.036 & 3.968 & 0.679 & 3.335 & 3.625 & 8.83 & B\\
        & 11 & 3.45509 & 2.293 & 0.630 & 0.022 & 0.002 & 0.8 & 0.15 & 2.006 & 3.058 & 3.98 & 0.67 & 3.321 & 3.596 & 8.78 & B\\
        & 12 & 2.51167 & 2.427 & 0.694 & 0.026 & 0.015 & 0.9 & 0.24 & 2.065 & 3.119 & 3.98 & 0.619 & 3.231 & 3.644 & 8.75 & B\\
        & 13 & 2.66580 & 2.437 & 0.717 & 0.017 & 0.011 & 1.1 & 0.17 & 2.176 & 2.726 & 4.016 & 0.676 & 3.327 & 3.651 & 8.78 & B\\
        & 14 & 2.94571 & 2.456 & 0.638 & 0.027 & 0.015 & 1.1 & 0.16 & 2.174 & 2.697 & 4.022 & 0.623 & 3.239 & 3.665 & 8.73 & B\\
        & 15 & 1.43394 & 1.986 & 0.837 & 0.023 & 0.027 & 1.2 & 0.03 & 2.171 & 3.136 & 3.921 & 0.6 & 3.184 & 3.666 & 9.28 & A\\
        & 16 & 2.55736 & 2.303 & 0.662 & 0.024 & 0.009 & 0.9 & 0.22 & 1.987 & 2.719 & 3.971 & 0.6 & 3.196 & 3.62 & 8.8 & B\\
        \multicolumn{3}{l}{TIC\,10756751\T\B} \\    
        & 1 & 1.23834 & 1.773 & 0.574 & 0.023 & 0.025 & 1.6 & 0.05 & 1.93 & 3.13 & 3.872 & 0.334 & 3.061 & 3.708 & 9.49 & A\\
        & 2 & 1.27960 & 1.980 & 0.674 & 0.014 & 0.027 & 1.7 & 0.03 & 2.185 & 3.093 & 3.918 & 0.411 & 3.279 & 3.721 & 9.37 & A\\
        & 3 & 1.26999 & 1.839 & 0.766 & 0.013 & 0.025 & 0.9 & 0.06 & 2.053 & 3.387 & 3.874 & 0.458 & 3.397 & 3.636 & 9.44 & A\\
        & 4 & 1.23910 & 2.060 & 0.813 & 0.012 & 0.023 & 1.6 & 0.09 & 2.24 & 2.884 & 3.944 & 0.483 & 3.457 & 3.704 & 9.28 & A\\
        & 5 & 1.53421 & 1.918 & 0.599 & 0.029 & 0.023 & 0.8 & 0.03 & 2.051 & 2.695 & 3.933 & 0.412 & 3.28 & 3.647 & 9.27 & A\\
        & 6 & 1.59119 & 1.898 & 0.566 & 0.025 & 0.028 & 0.6 & 0.05 & 1.983 & 2.65 & 3.925 & 0.396 & 3.239 & 3.657 & 9.33 & A\\
        & 7 & 1.29047 & 1.645 & 0.633 & 0.027 & 0.016 & 0.8 & 0.00 & 1.9 & 2.644 & 3.892 & 0.369 & 3.16 & 3.588 & 9.46 & A\\
        & 8 & 1.29674 & 1.066 & 1.044 & 0.025 & 0.024 & 0.8 & 0.00 & 1.668 & 2.757 & 3.822 & 0.433 & 3.334 & 3.564 & 9.91 & A\\
        & 9 & 1.20346 & 1.775 & 0.661 & 0.019 & 0.012 & 1.1 & 0.04 & 2.015 & 2.608 & 3.909 & 0.356 & 3.123 & 3.628 & 9.42 & A\\
        & 10 & 1.24970 & 1.655 & 0.747 & 0.028 & 0.028 & 0.9 & 0.08 & 1.856 & 3.795 & 3.82 & 0.437 & 3.345 & 3.612 & 9.54 & A\\
        \multicolumn{3}{l}{TIC\,37817410\T\B} \\ 
        & 1 & 1.47500 & 1.093 & 0.911 & 0.025 & 0.006 & 1.4 & 0.03 & 1.613 & 1.878 & 3.88 & 0.363 & 2.587 & 3.616 & 9.84 & B\\
        & 2 & 1.26929 & 1.835 & 0.531 & 0.023 & 0.027 & 1.5 & 0.29 & 1.594 & 1.916 & 3.89 & 0.327 & 2.505 & 3.734 & 9.51 & A\\
        & 3 & 1.25630 & 1.106 & 0.899 & 0.024 & 0.009 & 1.2 & 0.06 & 1.588 & 1.788 & 3.882 & 0.367 & 2.597 & 3.602 & 9.83 & B\\
        & 4 & 1.41540 & 1.827 & 0.503 & 0.020 & 0.011 & 0.5 & 0.27 & 1.578 & 1.573 & 3.872 & 0.35 & 2.567 & 3.583 & 9.11 & B\\
        & 5 & 1.43283 & 1.102 & 0.848 & 0.017 & 0.002 & 0.7 & 0.02 & 1.569 & 1.706 & 3.875 & 0.36 & 2.579 & 3.563 & 9.89 & B\\
        & 6 & 1.23132 & 1.621 & 0.555 & 0.026 & 0.028 & 1.4 & 0.20 & 1.562 & 1.925 & 3.877 & 0.353 & 2.567 & 3.69 & 9.6 & A\\
        & 7 & 1.39078 & 1.096 & 0.830 & 0.012 & 0.007 & 1.2 & 0.04 & 1.551 & 1.567 & 3.89 & 0.339 & 2.527 & 3.624 & 9.89 & A\\
        & 8 & 1.94405 & 1.057 & 0.798 & 0.012 & 0.014 & 1.8 & 0.00 & 1.543 & 1.55 & 3.89 & 0.306 & 2.442 & 3.667 & 9.95 & A\\
        & 9 & 1.23281 & 1.137 & 0.842 & 0.014 & 0.019 & 1.3 & 0.12 & 1.527 & 1.528 & 3.88 & 0.353 & 2.564 & 3.626 & 9.86 & B\\
        & 10 & 1.29737 & 1.120 & 0.830 & 0.026 & 0.005 & 0.9 & 0.12 & 1.516 & 1.718 & 3.871 & 0.335 & 2.517 & 3.574 & 9.8 & B\\
        & 11 & 2.51667 & 1.000 & 0.905 & 0.016 & 0.011 & 1.6 & 0.07 & 1.507 & 1.904 & 3.843 & 0.346 & 2.544 & 3.643 & 10.05 & A\\
        & 12 & 1.21457 & 1.639 & 0.610 & 0.023 & 0.019 & 0.8 & 0.29 & 1.507 & 1.846 & 3.858 & 0.367 & 2.598 & 3.607 & 9.57 & A\\
        & 13 & 1.37117 & 1.115 & 0.747 & 0.027 & 0.005 & 1.4 & 0.02 & 1.505 & 1.549 & 3.883 & 0.336 & 2.522 & 3.614 & 9.79 & B\\
        \multicolumn{3}{l}{TIC\,64437380\T\B} \\ 
        & 1 & 1.30566 & 1.024 & 0.916 & 0.019 & 0.013 & 1.1 & 0.08 & 1.608 & 2.729 & 3.806 & 0.263 & 2.167 & 3.612 & 10.01 & A\\
        & 2 & 1.26533 & 0.979 & 0.871 & 0.020 & 0.019 & 0.5 & 0.15 & 1.457 & 2.334 & 3.797 & 0.28 & 2.214 & 3.531 & 10.1 & A\\
        & 3 & 1.32244 & 0.989 & 0.830 & 0.029 & 0.011 & 0.7 & 0.10 & 1.466 & 2.178 & 3.82 & 0.275 & 2.199 & 3.553 & 10.0 & B\\
        & 4 & 2.35677 & 0.981 & 0.866 & 0.014 & 0.020 & 1.3 & 0.14 & 1.484 & 1.962 & 3.835 & 0.255 & 2.142 & 3.641 & 10.11 & B\\
        & 5 & 1.31652 & 0.993 & 0.706 & 0.012 & 0.014 & 0.8 & 0.06 & 1.427 & 1.651 & 3.856 & 0.218 & 2.034 & 3.606 & 10.09 & B\\
        & 6 & 1.70143 & 1.046 & 0.725 & 0.022 & 0.005 & 1.6 & 0.28 & 1.303 & 1.485 & 3.828 & 0.237 & 2.089 & 3.651 & 9.95 & B\\
        \multicolumn{3}{l}{TIC\,126945917\T\B} \\ 
        & 1 & 1.28362 & 1.092 & 0.823 & 0.024 & 0.003 & 0.6 & 0.09 & 1.534 & 2.001 & 3.845 & 0.304 & 1.57 & 3.551 & 9.88 & B\\
        & 2 & 1.38418 & 1.022 & 0.728 & 0.023 & 0.019 & 1.6 & 0.21 & 1.329 & 1.488 & 3.836 & 0.256 & 1.481 & 3.647 & 9.98 & B\\
        & 3 & 1.31652 & 0.993 & 0.706 & 0.012 & 0.014 & 0.8 & 0.06 & 1.337 & 1.427 & 3.839 & 0.316 & 1.59 & 3.596 & 10.06 & B\\
        & 4 & 2.62958 & 0.970 & 0.720 & 0.016 & 0.004 & 1.4 & 0.30 & 1.235 & 1.426 & 3.805 & 0.229 & 1.425 & 3.662 & 10.11 & B\\
        & 5 & 1.39068 & 1.041 & 0.555 & 0.030 & 0.009 & 1.7 & 0.12 & 1.264 & 1.387 & 3.834 & 0.231 & 1.43 & 3.653 & 9.89 & B\\
        & 6 & 1.27810 & 1.081 & 0.665 & 0.011 & 0.021 & 1.7 & 0.23 & 1.293 & 1.359 & 3.835 & 0.26 & 1.488 & 3.675 & 9.92 & B\\
        & 7 & 1.31712 & 1.131 & 0.562 & 0.018 & 0.010 & 1.5 & 0.24 & 1.178 & 1.229 & 3.795 & 0.317 & 1.592 & 3.639 & 9.83 & B\\
        & 8 & 1.56653 & 1.048 & 0.525 & 0.024 & 0.020 & 1.6 & 0.25 & 1.077 & 1.104 & 3.782 & 0.309 & 1.579 & 3.639 & 9.91 & B\\
        \hline
    \end{tabular}
\end{table*}

\begin{figure*}[hbt!]
\centering
\begin{subfigure}{0.49\textwidth}
  \centering
  \includegraphics[width=\linewidth]{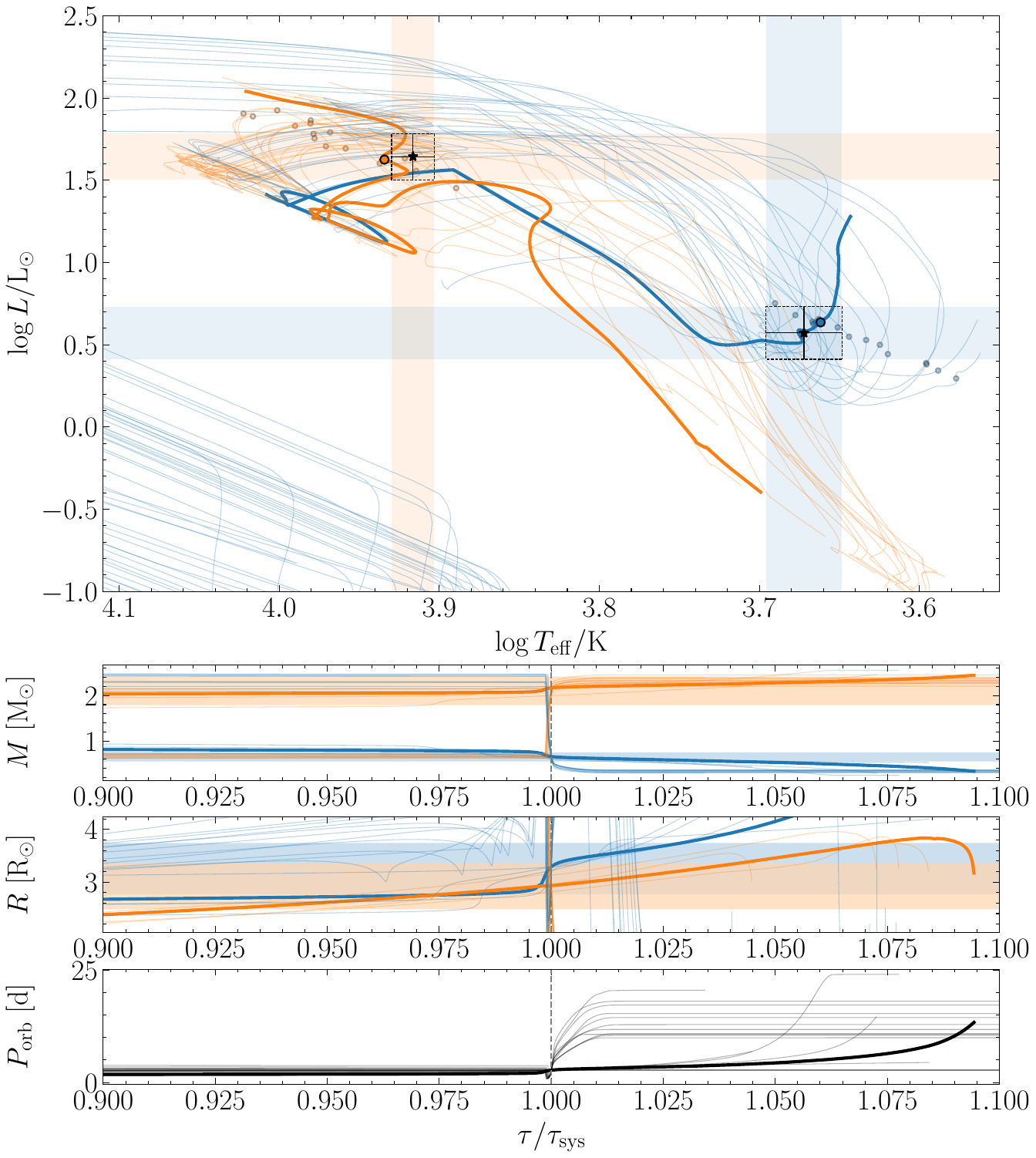}
  \caption{TIC~82474821}
\end{subfigure}
\hfill
\begin{subfigure}{0.49\textwidth}
  \centering
  \includegraphics[width=\linewidth]{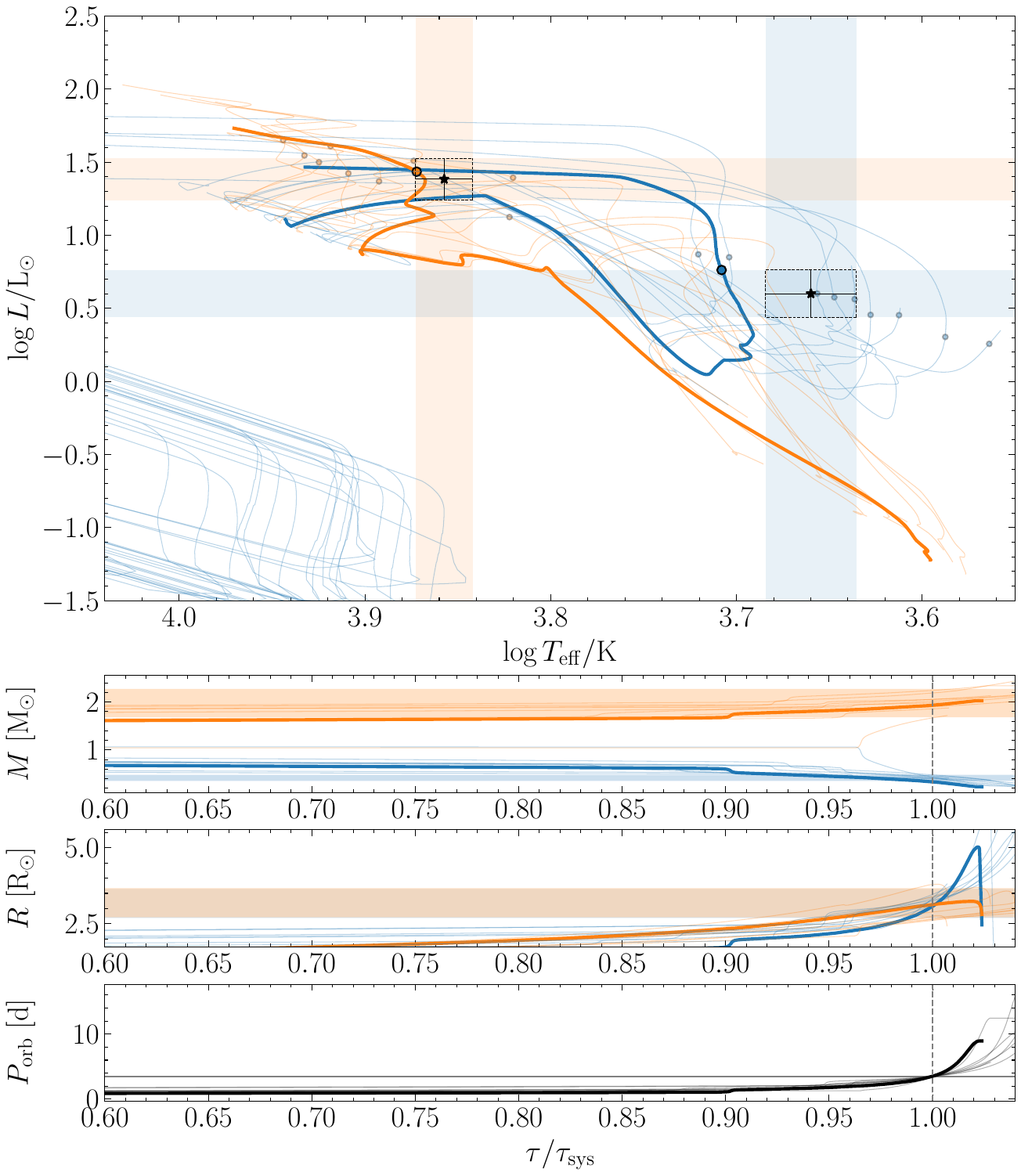}
  \caption{TIC~10756751}
\end{subfigure}\\
\vspace{0.5cm}
\begin{subfigure}{0.49\textwidth}
  \centering
  \includegraphics[width=\linewidth]{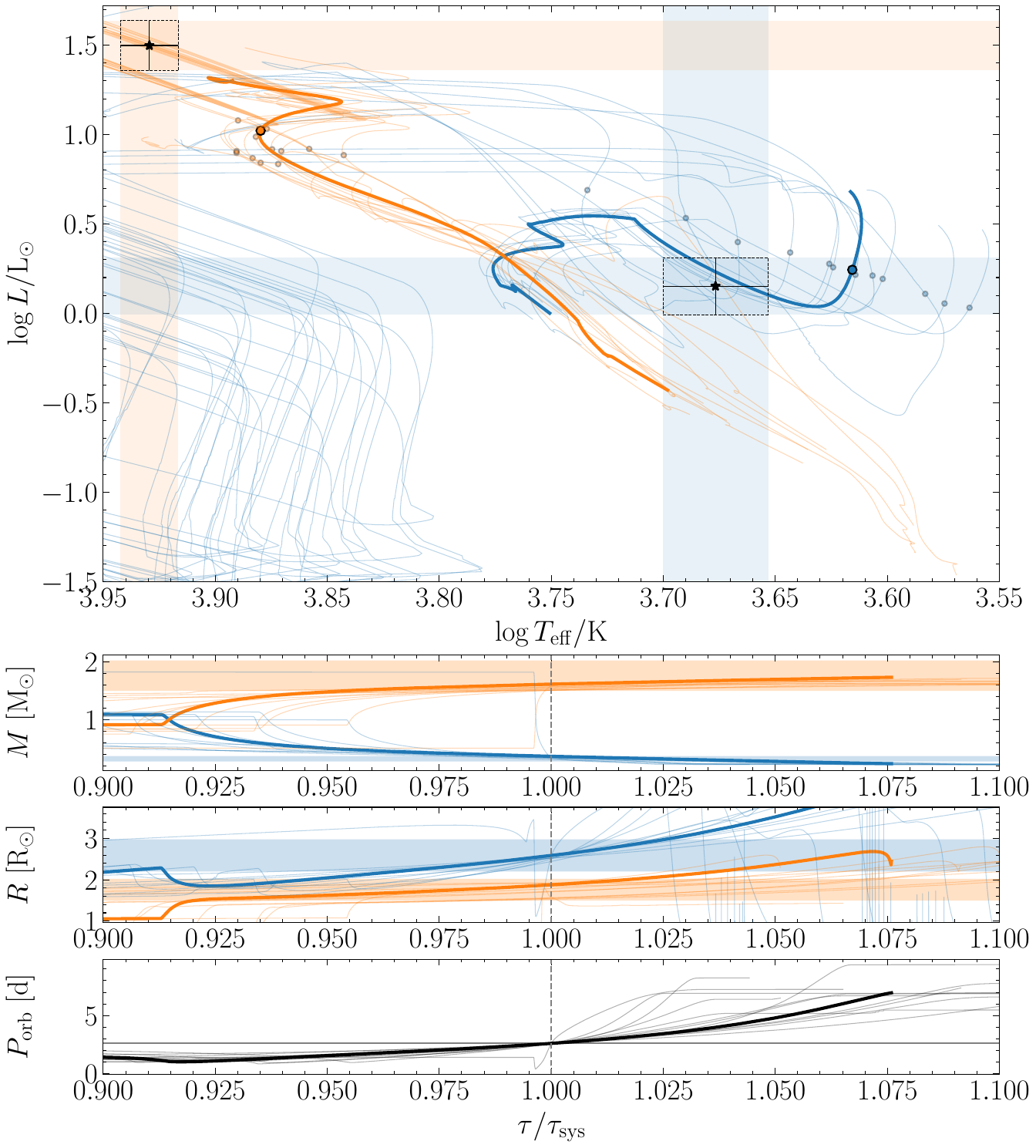}
  \caption{TIC~37817410}
\end{subfigure}
\hfill
\begin{subfigure}{0.49\textwidth}
  \centering
  \includegraphics[width=\linewidth]{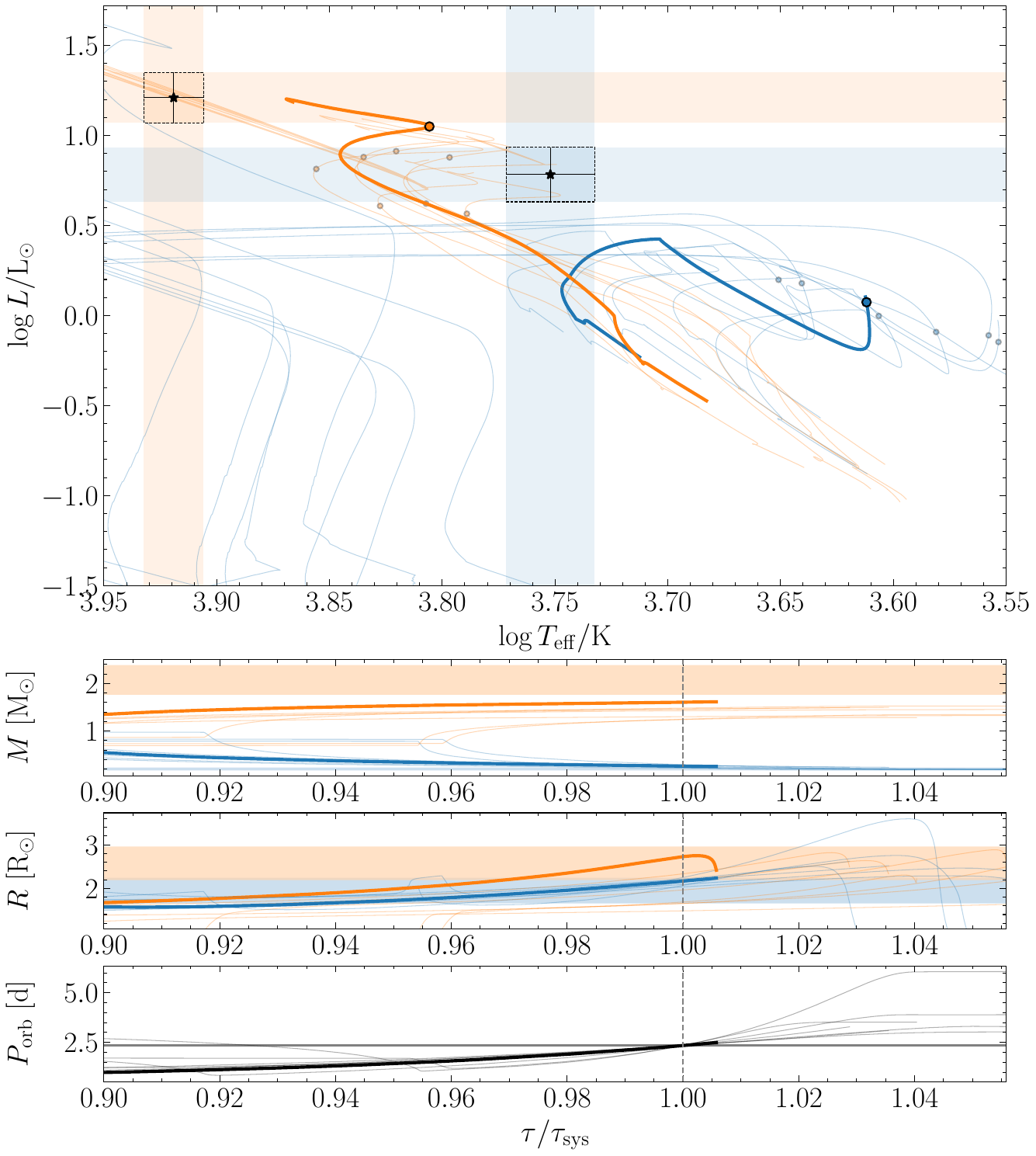}
  \caption{TIC~64437380}
\end{subfigure}
\end{figure*}

\begin{figure*}[hbt!]
\ContinuedFloat 
    \centering
\begin{subfigure}{0.49\textwidth}
  \centering
  \includegraphics[width=\linewidth]{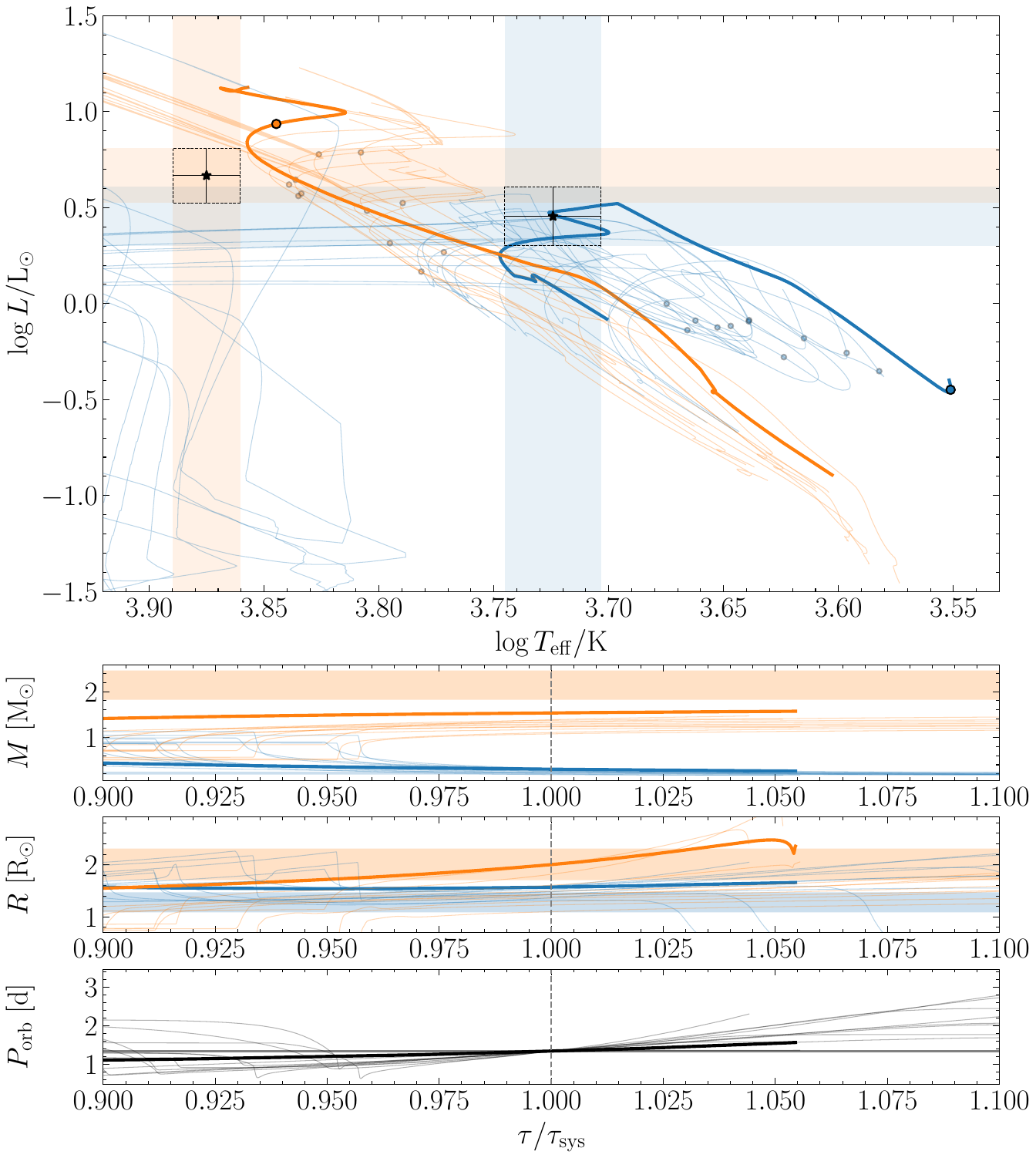}
  \caption{TIC~126945917}
\end{subfigure}

\caption{The HR diagrams and the evolution of masses, radii, and orbital period for the best-fit evolutionary models of the systems.}
\label{fig:MESA_plots}
\end{figure*}

\section{Pulsation Analysis}
To determine the pulsation frequencies for each system, we use the residuals left after subtracting the binary model from the LC. To ensure a good frequency resolution we recalculated binary models for the multiple TESS sectors obtaining the longest temporal baseline possible. The binary model is almost never perfect; as the LCs are affected due to spots and pulsations. To ensure that we have minimum contamination from the binary signal during the pulsation analysis we further clean the residuals in two steps. Firstly, we clean them for the first 100 harmonics of the orbital frequency. In the next step, we use the \textsc{Wotan} Python package to remove the long term trends arising due to imperfect subtraction of spot signals and eclipse ingress/egress profiles. In this step, we chopped off the LCs at the eclipses where the detrending was not sufficient.

\begin{figure*}[h!]
\centering
\begin{subfigure}{0.49\textwidth}
  \centering
  \includegraphics[width=\linewidth]{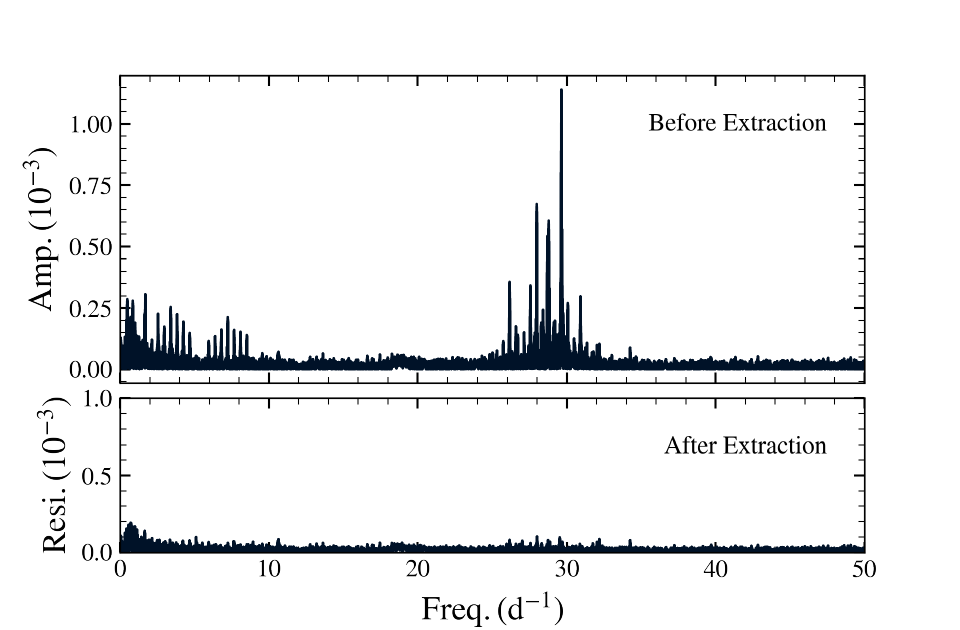}
  \caption{TIC~64437380}
\end{subfigure}
\hfill
\begin{subfigure}{0.49\textwidth}
  \centering
  \includegraphics[width=\linewidth]{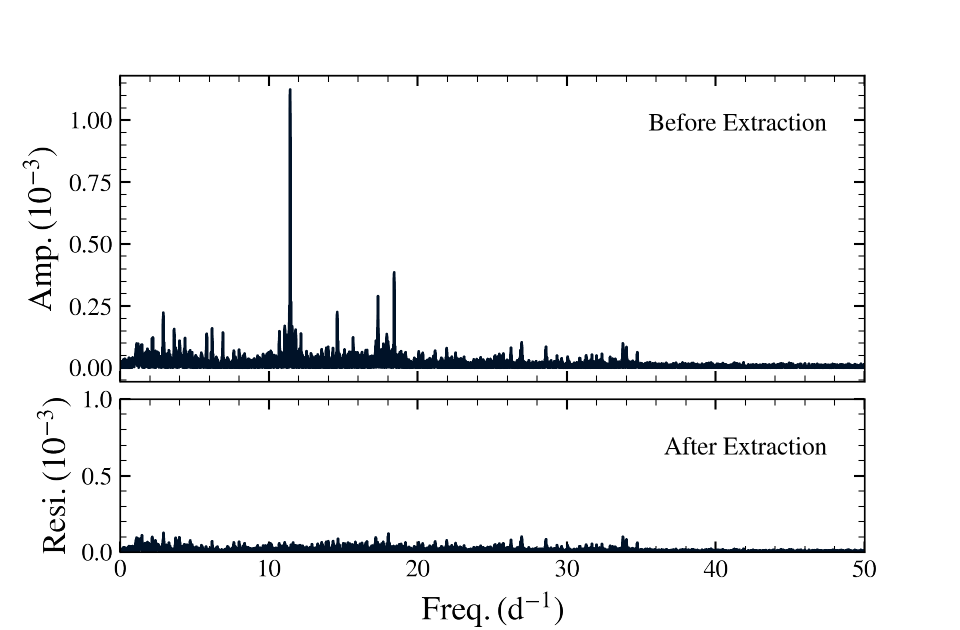}
  \caption{TIC~82474821}
\end{subfigure}
\centering
\begin{subfigure}{0.49\textwidth}
  \centering
  \includegraphics[width=\linewidth]{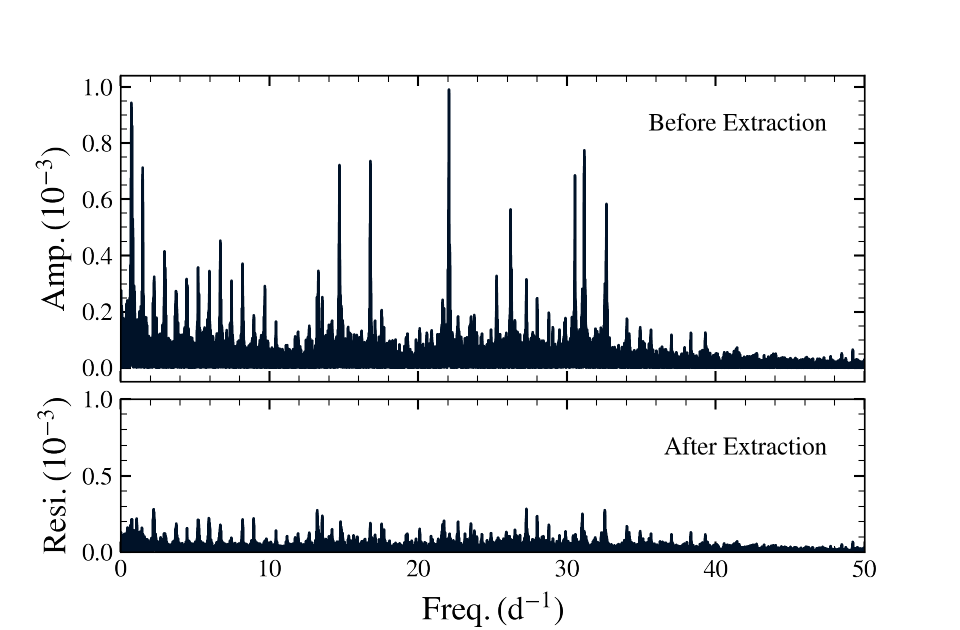}
  \caption{TIC~126945917}
\end{subfigure}
\hfill
\begin{subfigure}{0.49\textwidth}
  \centering
  \includegraphics[width=\linewidth]{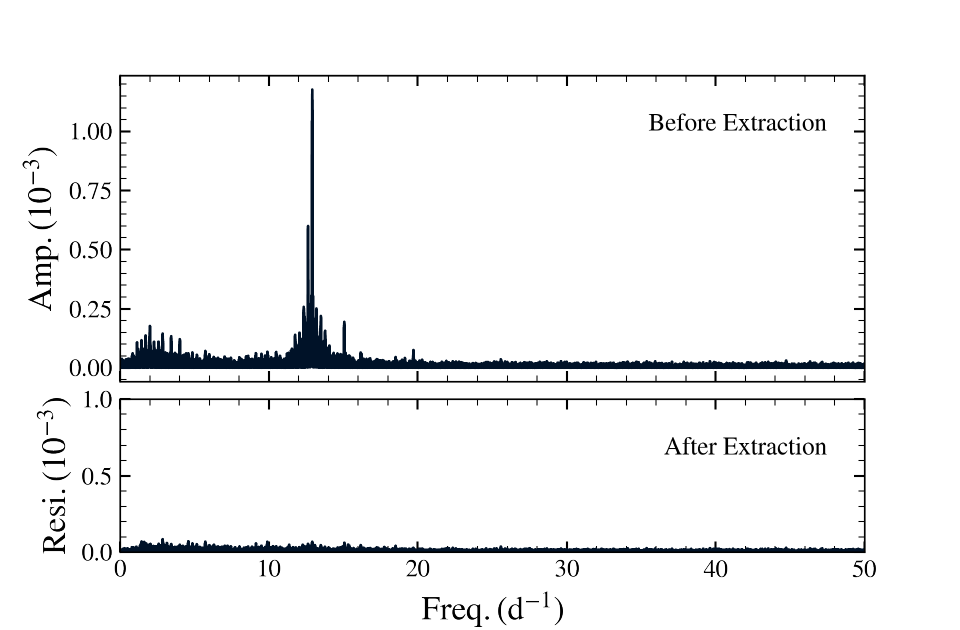}
  \caption{TIC~10756751}
\end{subfigure}
\hfill
\begin{subfigure}{0.49\textwidth}
  \centering
  \includegraphics[width=\linewidth]{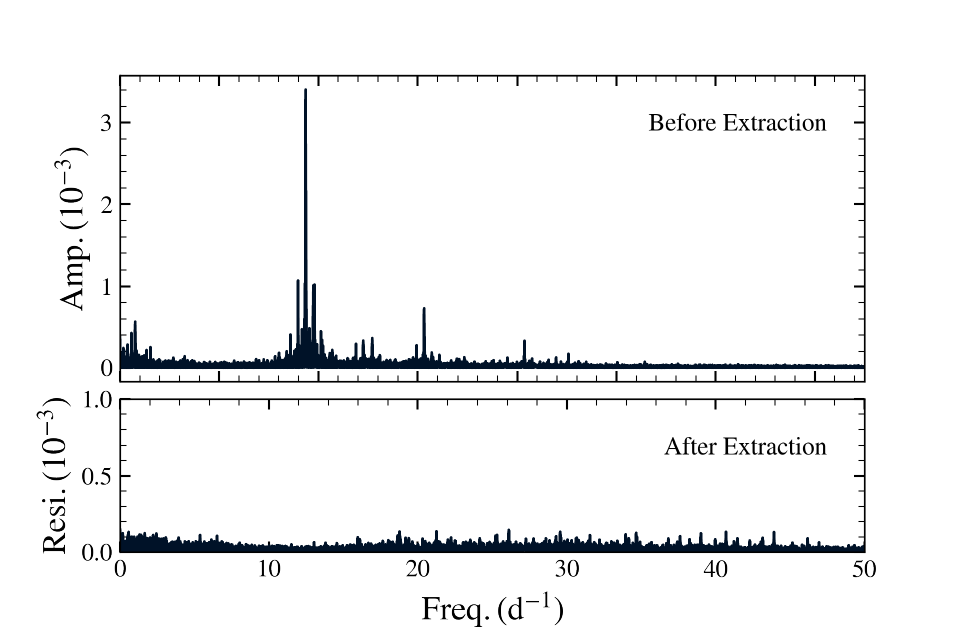}
  \caption{TIC~37817410}
\end{subfigure}

\caption{Periodograms for residuals obtained after subtracting binary models from the corresponding LCs.} 
\label{fig:periodograms}
\end{figure*}

To perform the time series analysis on these curated residuals we use v.1.2.9 of the \textsc{Period04}, based on \citet{2005CoAst.146...53L}. We perform the search for oscillation frequencies from 0 to 80 d$^{-1}$ to search for $\delta$ Scuti-$\gamma$ Doradus frequencies. A pre-whitening procedure was applied to extract all significant frequencies. The noise level was computed within a 2 d$^{-1}$ window around each peak, and the signal-to-noise ratio (SNR) was defined as the ratio of the peak amplitude to this average noise amplitude. Only frequencies with an SNR of 5 or greater after pre-whitening were considered significant \citep{2015MNRAS.448L..16B,2021A&A...656A.158B}. All the significant frequencies were then simultaneously fitted using sinusoids in order to obtain their amplitudes and phases. We show the original periodogram for this curated data and after extraction of all the significant frequencies on Figure~\ref{fig:periodograms}. A list of all the extracted frequencies, with their analytical errors \citep{1999DSSN...13...28M}, for each of the system is provided in Appendix \ref{appendix:B}.

\section{Conclusions}
In this study, we have presented a detailed analysis of five oEA systems selected from the CR\'{E}ME survey. By combining high-precision TESS photometry with multi-epoch high-resolution spectroscopy, we have derived precise LC and RV solutions that yield precise measurements of the orbital configuration of these binary systems and the fundamental stellar parameters.

MESA binary evolution simulations reveal that these systems have undergone either case A or case B mass transfer -- a conclusion supported by the observed low mass ratios. These findings underscore the significant role that binary interactions and mass transfer play in shaping the evolutionary pathways of intermediate-mass stars.


However, we note that the adopted \textsc{mesa-binary} models for our systems do not fully reproduce the observed stellar properties. This discrepancy may arise from various factors, including the treatment of convective overshooting, mass transfer efficiency, angular momentum loss, and stellar rotation. Further refinements are required to better align theoretical predictions with observations, which is crucial for accurately determining the progenitor properties and improving our understanding of the mass transfer process.

Our results contribute to the growing catalog of well-characterised oscillating EBs, providing essential benchmarks for testing and refining theoretical models of stellar structure and evolution. We aim to extend the asteroseismic analysis of these targets in a future study, specifically mode-identification using multi-colour photometry. Seismic modelling will help us further constrain the possible formation scenarios and provide insights on the changes in the internal structure and composition of the stars before and after the mass transfer process and the effect of such mass transfer events on the pulsation properties of the progenitors.


\section*{Acknowledgements}

We acknowledge support provided by the Polish National Science Center through grants no. \mbox{2017/27/B/ST9/02727}, \mbox{2021/41/N/ST9/02746}, \mbox{2021/43/B/ST9/02972}, \mbox{2023/49/B/ST9/01671}, and \mbox{2024/53/N/ST9/03885}. A. Moharana acknowledges support from the UK Science and Technology Facilities Council (STFC) under grant number ST/Y002563/1.

This research made use of data collected at ESO under programmes 088.D-0080, 089.D-0097, 090.D-0061, 091.D-0145, 094.A-9029, as well as through CNTAC proposal 2014B-067.

This research is based in part on data collected at Subaru Telescope, which is operated by the National Astronomical Observatory of Japan. We are honored and grateful for the opportunity of observing the Universe from Maunakea, which has the cultural, historical and natural significance in Hawaii.

This research made use of Lightkurve, a Python package for \textit{Kepler} and TESS data analysis \citep{2018ascl.soft12013L}. This research uses data collected by the TESS mission (Guest Investigator proposals: G011083, G011154, G03028, G03251, G04047, G04171, G04234, G05003, G05078, G05125, G06057) which are publicly available from the MAST portal. Funding for the TESS mission is provided by NASA’s Science Mission directorate.

%
%


\bibliographystyle{aa}
\bibliography{pawar}


\appendix

\section{Eclipse Minima Times}
\label{appendix:A}

\begin{table*}[h!]
\centering
\small
\caption{Times of primary (integer cycle) and secondary (half-integer cycle) eclipse minima for the target TIC~64437380}
\begin{tabular}{ccc @{\hspace{1cm}} ccc @{\hspace{1cm}} ccc}
\hline
BJD-2457000 & Cycle & 1-$\sigma$ & BJD-2457000 & Cycle & 1-$\sigma$ & BJD-2457000 & Cycle & 1-$\sigma$ \\
\hline
2362.396193 & -301.5 & 0.000364 & 2382.344695 & -293.0 & 0.000090 & 3079.377001 & 4.0   & 0.000099 \\
2363.569631 & -301.0 & 0.000099 & 2383.519211 & -292.5 & 0.000385 & 3080.550676 & 4.5   & 0.000328 \\
2364.743176 & -300.5 & 0.000342 & 2384.691686 & -292.0 & 0.000096 & 3081.723891 & 5.0   & 0.000099 \\
2365.916519 & -300.0 & 0.000091 & 2385.866477 & -291.5 & 0.000369 & 3082.897978 & 5.5   & 0.000405 \\
2367.090087 & -299.5 & 0.000302 & 2387.038335 & -291.0 & 0.000095 & 3084.070877 & 6.0   & 0.000102 \\
2368.263561 & -299.0 & 0.000097 & 2388.212779 & -290.5 & 0.000338 & 3085.244844 & 6.5   & 0.000358 \\
2369.437316 & -298.5 & 0.000343 & 2389.385625 & -290.0 & 0.000101 & 3086.417780 & 7.0   & 0.000097 \\
2370.610299 & -298.0 & 0.000098 & 3068.814548 & -0.5   & 0.000394 & 3087.592269 & 7.5   & 0.000366 \\
2371.784380 & -297.5 & 0.000345 & 3069.989692 &  0.0   & 0.000104 & 3088.764684 & 8.0   & 0.000100 \\
2372.957192 & -297.0 & 0.000093 & 3071.163243 &  0.5   & 0.000257 & 3089.938724 & 8.5   & 0.000367 \\
2374.131110 & -296.5 & 0.000283 & 3072.336533 &  1.0   & 0.000098 & 3091.111695 & 9.0   & 0.000098 \\
2376.477589 & -295.5 & 0.000369 & 3073.509847 &  1.5   & 0.000394 & 3092.286170 & 9.5   & 0.000403 \\
2377.650928 & -295.0 & 0.000089 & 3074.683356 &  2.0   & 0.000101 & 3093.458533 & 10.0  & 0.000099 \\
2378.825263 & -294.5 & 0.000266 & 3075.857370 &  2.5   & 0.000378 & 3094.632899 & 10.5  & 0.000370 \\
2379.997904 & -294.0 & 0.000090 & 3077.030191 &  3.0   & 0.000101 & 3095.805511 & 11.0  & 0.000107 \\
2381.171891 & -293.5 & 0.000322 & 3078.203905 &  3.5   & 0.000371 &             &       &          \\
\hline
\end{tabular}
\end{table*}

\begin{table*}[h!]
\centering
\small
\caption{Times of primary (integer cycle) and secondary (half-integer cycle) eclipse minima for the target TIC~82474821}
\begin{tabular}{ccc@{\hspace{1cm}}ccc@{\hspace{1cm}}ccc}
\hline
BJD--2457000 & Cycle  & 1-$\sigma$ & BJD--2457000 & Cycle  & 1-$\sigma$ & BJD--2457000 & Cycle  & 1-$\sigma$ \\
\hline
2229.519557 & -277.0 & 0.000071 & 2266.613007 & -263.5 & 0.003327 & 2983.668094 &  -2.5 & 0.000431 \\
2230.897484 & -276.5 & 0.000353 & 2268.507459 & -263.0 & 0.182495 & 2985.037314 &  -2.0 & 0.000064 \\
2232.266973 & -276.0 & 0.000072 & 2269.362539 & -262.5 & 0.003154 & 2986.415628 &  -1.5 & 0.000320 \\
2233.644984 & -275.5 & 0.000341 & 2270.729715 & -262.0 & 0.000742 & 2987.784656 &  -1.0 & 0.000070 \\
2235.014119 & -275.0 & 0.000073 & 2272.094526 & -261.5 & 0.002129 & 2989.162642 &  -0.5 & 0.000312 \\
2236.392158 & -274.5 & 0.000329 & 2273.477106 & -261.0 & 0.000846 & 2990.532038 &   0.0 & 0.000073 \\
2237.761462 & -274.0 & 0.000077 & 2274.849345 & -260.5 & 0.002542 & 2991.909698 &   0.5 & 0.000360 \\
2239.138815 & -273.5 & 0.000518 & 2276.224417 & -260.0 & 0.000761 & 2993.279487 &   1.0 & 0.000067 \\
2240.508938 & -273.0 & 0.000072 & 2277.598701 & -259.5 & 0.001996 & 2994.656550 &   1.5 & 0.000330 \\
2243.256098 & -272.0 & 0.000076 & 2278.971788 & -259.0 & 0.000731 & 2996.026867 &   2.0 & 0.000068 \\
2244.633544 & -271.5 & 0.000351 & 2963.058400 &  -10.0 & 0.000092 & 2997.403772 &   2.5 & 0.000353 \\
2246.003488 & -271.0 & 0.000070 & 2964.437221 &   -9.5 & 0.000383 & 2998.774203 &   3.0 & 0.000065 \\
2247.380696 & -270.5 & 0.000807 & 2965.805988 &   -9.0 & 0.000072 & 3000.150676 &   3.5 & 0.000312 \\
2248.750723 & -270.0 & 0.000071 & 2967.184534 &   -8.5 & 0.000256 & 3001.521649 &   4.0 & 0.000069 \\
2250.128238 & -269.5 & 0.000411 & 2968.553322 &   -8.0 & 0.000072 & 3002.898341 &   4.5 & 0.000425 \\
2251.498224 & -269.0 & 0.000078 & 2969.931891 &   -7.5 & 0.000428 & 3004.268991 &   5.0 & 0.000067 \\
2252.875028 & -268.5 & 0.000462 & 2971.300504 &   -7.0 & 0.000068 & 3005.645087 &   5.5 & 0.000411 \\
2255.622050 & -267.5 & 0.001273 & 2972.679335 &   -6.5 & 0.000349 & 3007.016419 &   6.0 & 0.000069 \\
2256.993027 & -267.0 & 0.000583 & 2974.048015 &   -6.0 & 0.000067 & 3008.391846 &   6.5 & 0.000379 \\
2258.369864 & -266.5 & 0.002765 & 2975.425040 &   -5.5 & 0.000607 & 3009.763798 &   7.0 & 0.000066 \\
2259.740416 & -266.0 & 0.000649 & 2976.795260 &   -5.0 & 0.000078 & 3011.138809 &   7.5 & 0.000578 \\
2261.115217 & -265.5 & 0.003119 & 2978.173736 &   -4.5 & 0.000303 & 3012.511181 &   8.0 & 0.000067 \\
2262.487751 & -265.0 & 0.000754 & 2979.542649 &   -4.0 & 0.000069 & 3013.885406 &   8.5 & 0.000427 \\
2263.864059 & -264.5 & 0.003889 & 2980.920741 &   -3.5 & 0.000495 &             &       &         \\
2265.235091 & -264.0 & 0.000622 & 2982.289848 &   -3.0 & 0.000102 &             &       &         \\
\hline
\end{tabular}
\end{table*}

\begin{table*}[h!]
\centering
\small
\caption{Times of primary (integer cycle) and secondary (half-integer cycle) eclipse minima for the target TIC~126945917}
\begin{tabular}{ccc@{\hspace{1cm}}ccc@{\hspace{1cm}}ccc}
\hline
BJD--2457000 & Cycle & 1-$\sigma$ & BJD--2457000 & Cycle & 1-$\sigma$ & BJD--2457000 & Cycle & 1-$\sigma$ \\
\hline
1325.624401 & -530.5  & 0.000451 & 2039.745243 & 2.0   & 0.000095 & 2075.283022 & 28.5  & 0.000238 \\
1326.291243 & -530.0  & 0.000103 & 2040.416943 & 2.5   & 0.000296 & 2075.954571 & 29.0  & 0.000094 \\
1326.965744 & -529.5  & 0.000288 & 2041.086262 & 3.0   & 0.000093 & 2076.623998 & 29.5  & 0.000359 \\
1327.632305 & -529.0  & 0.000084 & 2041.758065 & 3.5   & 0.000317 & 2077.295631 & 30.0  & 0.000107 \\
1328.306375 & -528.5  & 0.000336 & 2042.427350 & 4.0   & 0.000092 & 2077.965481 & 30.5  & 0.000275 \\
1328.973460 & -528.0  & 0.000088 & 2043.099297 & 4.5   & 0.000317 & 2078.636687 & 31.0  & 0.000083 \\
1329.647835 & -527.5  & 0.000329 & 2043.768235 & 5.0   & 0.000095 & 2079.306818 & 31.5  & 0.000309 \\
1330.314408 & -527.0  & 0.000093 & 2044.440086 & 5.5   & 0.000364 & 2079.977753 & 32.0  & 0.000086 \\
1330.988217 & -526.5  & 0.000329 & 2045.109516 & 6.0   & 0.000085 & 2080.648042 & 32.5  & 0.000336 \\
1331.655613 & -526.0  & 0.000091 & 2045.781201 & 6.5   & 0.000346 & 2081.318806 & 33.0  & 0.000080 \\
1332.329617 & -525.5  & 0.000285 & 2046.450564 & 7.0   & 0.000093 & 2081.988969 & 33.5  & 0.000248 \\
1332.996506 & -525.0  & 0.000084 & 2047.121998 & 7.5   & 0.000274 & 2082.659827 & 34.0  & 0.000091 \\
1333.670803 & -524.5  & 0.000277 & 2047.791704 & 8.0   & 0.000096 & 2083.330184 & 34.5  & 0.000271 \\
1334.337687 & -524.0  & 0.000091 & 2049.132789 & 9.0   & 0.000119 & 2084.000925 & 35.0  & 0.000083 \\
1335.012311 & -523.5  & 0.000327 & 2049.804327 & 9.5   & 0.000232 & 3154.853333 & 833.5 & 0.000375 \\
1335.678640 & -523.0  & 0.000094 & 2050.473989 & 10.0  & 0.000096 & 3155.521829 & 834.0 & 0.000090 \\
1336.352980 & -522.5  & 0.000457 & 2051.145023 & 10.5  & 0.000315 & 3156.194623 & 834.5 & 0.000294 \\
1337.019604 & -522.0  & 0.000088 & 2051.814959 & 11.0  & 0.000081 & 3156.862943 & 835.0 & 0.000085 \\
1337.693816 & -521.5  & 0.000291 & 2052.486221 & 11.5  & 0.000254 & 3157.535633 & 835.5 & 0.000269 \\
1338.360713 & -521.0  & 0.000094 & 2053.156252 & 12.0  & 0.000097 & 3158.203988 & 836.0 & 0.000089 \\
1339.701580 & -520.0  & 0.000114 & 2053.826849 & 12.5  & 0.000339 & 3158.876918 & 836.5 & 0.000232 \\
1340.375732 & -519.5  & 0.000197 & 2054.497252 & 13.0  & 0.000097 & 3159.545041 & 837.0 & 0.000087 \\
1341.042948 & -519.0  & 0.000095 & 2055.168208 & 13.5  & 0.000296 & 3160.218065 & 837.5 & 0.000296 \\
1341.717207 & -518.5  & 0.000240 & 2055.838296 & 14.0  & 0.000092 & 3160.886119 & 838.0 & 0.000088 \\
1342.383888 & -518.0  & 0.000086 & 2056.508738 & 14.5  & 0.000427 & 3161.558909 & 838.5 & 0.000253 \\
1343.057811 & -517.5  & 0.000385 & 2057.179431 & 15.0  & 0.000098 & 3162.227088 & 839.0 & 0.000096 \\
1343.724960 & -517.0  & 0.000094 & 2057.850175 & 15.5  & 0.000258 & 3162.900133 & 839.5 & 0.000333 \\
1344.399294 & -516.5  & 0.000442 & 2058.520485 & 16.0  & 0.000089 & 3163.568231 & 840.0 & 0.000092 \\
1345.066070 & -516.0  & 0.000086 & 2059.190904 & 16.5  & 0.000329 & 3164.241169 & 840.5 & 0.000348 \\
1345.740220 & -515.5  & 0.000340 & 2059.861664 & 17.0  & 0.000085 & 3164.909259 & 841.0 & 0.000093 \\
1346.407152 & -515.0  & 0.000098 & 2060.531989 & 17.5  & 0.000230 & 3168.932336 & 844.0 & 0.000084 \\
1347.081189 & -514.5  & 0.000337 & 2061.870880 & 18.5  & 0.000446 & 3169.605456 & 844.5 & 0.000303 \\
1347.748468 & -514.0  & 0.000092 & 2062.543737 & 19.0  & 0.000095 & 3170.273441 & 845.0 & 0.000088 \\
1348.422002 & -513.5  & 0.001262 & 2063.213993 & 19.5  & 0.000363 & 3170.946689 & 845.5 & 0.000363 \\
1349.089074 & -513.0  & 0.000235 & 2063.884917 & 20.0  & 0.000084 & 3171.614611 & 846.0 & 0.000090 \\
1349.763999 & -512.5  & 0.000255 & 2064.554848 & 20.5  & 0.000297 & 3172.287721 & 846.5 & 0.000346 \\
1350.430374 & -512.0  & 0.000084 & 2065.226050 & 21.0  & 0.000088 & 3172.955643 & 847.0 & 0.000088 \\
1351.104822 & -511.5  & 0.000341 & 2065.896113 & 21.5  & 0.000247 & 3173.628501 & 847.5 & 0.000242 \\
1351.771507 & -511.0  & 0.000091 & 2066.567021 & 22.0  & 0.000084 & 3174.296724 & 848.0 & 0.000087 \\
1352.445971 & -510.5  & 0.000256 & 2067.236661 & 22.5  & 0.000252 & 3174.969441 & 848.5 & 0.000252 \\
1353.112493 & -510.0  & 0.000133 & 2067.908112 & 23.0  & 0.000094 & 3175.637867 & 849.0 & 0.000094 \\
2036.394339 & -0.5    & 0.000242 & 2068.577663 & 23.5  & 0.000203 & 3176.310933 & 849.5 & 0.000257 \\
2037.063025 & 0.0     & 0.000091 & 2069.249207 & 24.0  & 0.000089 & 3176.978922 & 850.0 & 0.000093 \\
2037.735419 & 0.5     & 0.000292 & 2069.918770 & 24.5  & 0.000408 & 3177.652184 & 850.5 & 0.000276 \\
2038.404083 & 1.0     & 0.000092 & 2070.590238 & 25.0  & 0.000087 & 3178.319982 & 851.0 & 0.000089 \\
2039.076336 & 1.5     & 0.000310 & 2071.259535 & 25.5  & 0.000419 & 3178.993127 & 851.5 & 0.000274 \\
\hline
\end{tabular}
\end{table*}

\begin{table*}[h!]
\centering
\small
\caption{Times of primary (integer cycle) and secondary (half-integer cycle) eclipse minima for the target TIC~10756751}
\begin{tabular}{ccc@{\hspace{2em}}ccc@{\hspace{2em}}ccc}
\hline
BJD--2457000 & Cycle  & 1-$\sigma$ & BJD--2457000 & Cycle  & 1-$\sigma$ & BJD--2457000 & Cycle  & 1-$\sigma$ \\
\hline
1386.258438 & -522.5  & 0.000554 & 2462.440775 & -214.0  & 0.000107 & 2495.584159 & -204.5  & 0.000524 \\
1387.997561 & -522.0  & 0.000109 & 2464.188013 & -213.5  & 0.000459 & 2497.325205 & -204.0  & 0.000108 \\
1389.747137 & -521.5  & 0.000491 & 2465.929209 & -213.0  & 0.000108 & 3208.967778 & 0.0     & 0.000112 \\
1391.485993 & -521.0  & 0.000101 & 2467.677443 & -212.5  & 0.000429 & 3210.711594 & 0.5     & 0.000502 \\
1393.235962 & -520.5  & 0.000559 & 2469.417663 & -212.0  & 0.000110 & 3212.455909 & 1.0     & 0.000100 \\
1396.723826 & -519.5  & 0.000647 & 2474.653491 & -210.5  & 0.000495 & 3214.201061 & 1.5     & 0.000523 \\
1398.463186 & -519.0  & 0.000107 & 2476.394541 & -210.0  & 0.000103 & 3215.944313 & 2.0     & 0.000107 \\
1400.212003 & -518.5  & 0.000599 & 2478.142102 & -209.5  & 0.000430 & 3217.689992 & 2.5     & 0.000603 \\
1401.951742 & -518.0  & 0.000106 & 2479.882923 & -209.0  & 0.000112 & 3222.921333 & 4.0     & 0.000101 \\
1403.701315 & -517.5  & 0.000474 & 2481.629650 & -208.5  & 0.000412 & 3224.667606 & 4.5     & 0.000454 \\
1405.440133 & -517.0  & 0.000106 & 2488.605930 & -206.5  & 0.000516 & 3226.409815 & 5.0     & 0.000115 \\
2448.486848 & -218.0  & 0.000103 & 2490.348526 & -206.0  & 0.000105 & 3228.155380 & 5.5     & 0.000562 \\
2450.234401 & -217.5  & 0.000542 & 2492.095454 & -205.5  & 0.000583 & 3229.898167 & 6.0     & 0.000115 \\
2451.975367 & -217.0  & 0.000104 & 2493.836936 & -205.0  & 0.000113 & 3231.644278 & 6.5     & 0.000484 \\
\hline
\end{tabular}
\end{table*}

\begin{table*}[h!]
\centering
\small
\caption{Times of primary (integer cycle) and secondary (half-integer cycle) eclipse minima for the target TIC~37817410}
\begin{tabular}{ccc@{\hspace{1cm}}ccc@{\hspace{1cm}}ccc}
\hline
BJD--2457000 & Cycle & 1-$\sigma$ & BJD--2457000 & Cycle & 1-$\sigma$ & BJD--2457000 & Cycle & 1-$\sigma$ \\
\hline
1438.007277 &  -0.5  & 0.001944 & 1457.552194 &   7.0  & 0.000072 & 2184.662999 & 286.0 & 0.000071 \\
1439.309131 &   0.0  & 0.000070 & 1458.855020 &   7.5  & 0.000242 & 2185.979140 & 286.5 & 0.000565 \\
1440.612396 &   0.5  & 0.001107 & 1460.158395 &   8.0  & 0.000070 & 2187.269224 & 287.0 & 0.000072 \\
1441.915237 &   1.0  & 0.000075 & 1461.461928 &   8.5  & 0.000564 & 2188.574951 & 287.5 & 0.000442 \\
1443.220008 &   1.5  & 0.000409 & 1462.764513 &   9.0  & 0.000078 & 2189.875199 & 288.0 & 0.000065 \\
1444.521329 &   2.0  & 0.000078 & 1464.067669 &   9.5  & 0.000549 & 2191.180115 & 288.5 & 0.000943 \\
1445.825641 &   2.5  & 0.000599 & 2174.238577 & 282.0  & 0.000090 & 2192.481387 & 289.0 & 0.000073 \\
1447.127503 &   3.0  & 0.000075 & 2175.544447 & 282.5  & 0.000361 & 2193.786243 & 289.5 & 0.000551 \\
1448.431964 &   3.5  & 0.000809 & 2176.844705 & 283.0  & 0.000073 & 2195.087532 & 290.0 & 0.000073 \\
1449.733627 &   4.0  & 0.000079 & 2178.150458 & 283.5  & 0.000652 & 2196.392149 & 290.5 & 0.000914 \\
1452.339858 &   5.0  & 0.000075 & 2179.450782 & 284.0  & 0.000077 & 2197.693696 & 291.0 & 0.000072 \\
1453.642769 &   5.5  & 0.000477 & 2180.756363 & 284.5  & 0.000566 & 2198.997590 & 291.5 & 0.000702 \\
1454.946031 &   6.0  & 0.000071 & 2182.056855 & 285.0  & 0.000069 & 2200.299252 & 292.0 & 0.000139 \\
1456.249326 &   6.5  & 0.000566 & 2183.362848 & 285.5  & 0.000321 &             &      &         \\
\hline
\end{tabular}
\end{table*}


\section{Pulsation Frequencies}
\label{appendix:B}
\begin{table*}[h!]
\centering
\small
\caption{Frequencies used to compute the Least-Squares Fourier fitting for the target TIC~64437380}

\begin{tabular}{lcccc}

Label & Frequency ($\mathrm{d^{-1}}$) & Amplitude & Phase \\
\hline
F1  &  29.631632 $\pm$ 9.28e-06     & 0.000997 $\pm$ 1.24e-05     & 0.700701   $\pm$  0.001965  \\
F2  &  27.973431 $\pm$ 1.30e-05     & 0.000712 $\pm$ 1.24e-05     & 0.039448   $\pm$  0.002769\\
F3  &  28.688072 $\pm$ 1.82e-05     & 0.000468 $\pm$ 1.24e-05     & 0.970814   $\pm$  0.003872\\
F4  &  28.776626 $\pm$ 2.54e-05     & 0.000360 $\pm$ 1.24e-05     & 0.090007  $\pm$   0.005397\\
F5  &  26.149574 $\pm$ 2.67e-05     & 0.000348 $\pm$ 1.24e-05     & 0.105014   $\pm$  0.005657\\
F6  &  1.7057163 $\pm$ 3.28e-05     & 0.000282 $\pm$ 1.24e-05     & 0.119956   $\pm$  0.006951\\
F7  &  0.4896069 $\pm$ 2.92e-05     & 0.000317 $\pm$ 1.24e-05     & 0.243341   $\pm$  0.006188\\
F8  &  29.662922 $\pm$ 3.17e-05     & 0.000291 $\pm$ 1.24e-05     & 0.661731   $\pm$  0.006718\\
F9  &  27.124015 $\pm$ 3.34e-05     & 0.000276 $\pm$ 1.24e-05     & 0.687824   $\pm$  0.007091\\
F10 &  3.4059107 $\pm$ 2.74e-05     & 0.000338 $\pm$ 1.24e-05     & 0.237469   $\pm$  0.005810\\
F11 &  4.2595528 $\pm$ 3.77e-05     & 0.000246 $\pm$ 1.24e-05     & 0.554921   $\pm$  0.007999\\
F12 &  2.5537002 $\pm$ 3.44e-05     & 0.000271 $\pm$ 1.24e-05     & 0.528809   $\pm$  0.007298\\
F13 &  0.8575960 $\pm$ 3.94e-05     & 0.000236 $\pm$ 1.24e-05     & 0.499437   $\pm$  0.008357\\
F14 &  7.2421192 $\pm$ 3.42e-05     & 0.000272 $\pm$ 1.24e-05     & 0.034559  $\pm$   0.007251\\
F15 &  0.6084296 $\pm$ 4.48e-05     & 0.000206 $\pm$ 1.24e-05     & 0.852651   $\pm$  0.009492\\
F16 &  30.917208 $\pm$ 4.65e-05     & 0.000163 $\pm$ 1.24e-05     & 0.427604   $\pm$  0.009849\\
F17 &  8.0957614 $\pm$ 4.48e-05     & 0.000207 $\pm$ 1.24e-05     & 0.350799   $\pm$  0.009497\\
F18 &  6.3913841 $\pm$ 5.47e-05     & 0.000170 $\pm$ 1.24e-05     & 0.746607   $\pm$  0.011592\\
F19 &  29.706199 $\pm$ 5.23e-05     & 0.000137 $\pm$ 1.24e-05     & 0.578408   $\pm$  0.011089\\
F20 &  29.557108 $\pm$ 6.48e-05     & 0.000129 $\pm$ 1.24e-05     & 0.798661   $\pm$  0.013720\\
F21 &  30.916514 $\pm$ 4.54e-05     & 0.000129 $\pm$ 1.24e-05     & 0.556667   $\pm$  0.009624\\
F22 &  28.690173 $\pm$ 2.36e-05     & 0.000129 $\pm$ 1.24e-05     & 0.528247   $\pm$  0.005002\\
\hline
\end{tabular}

\label{tab:TIC644_freqs}
\end{table*}

\begin{table*}[h!]
\centering
\small
\caption{Frequencies used to compute the Least-Squares Fourier fitting for the target TIC~82474821}

\begin{tabular}{lcccc}

Label & Frequency ($\mathrm{d^{-1}}$) & Amplitude & Phase \\
\hline
F1  &    11.438400 $\pm$ 4.33e-06   & 0.001147  $\pm$  6.95e-06     &  0.138650  $\pm$ 0.000948 \\
F2  &    18.428984 $\pm$ 1.28e-05   & 0.000393  $\pm$  6.95e-06     &  0.312123  $\pm$ 0.002815 \\
F3  &    17.336435 $\pm$ 1.72e-05   & 0.000293  $\pm$  6.95e-06     &  0.402974  $\pm$ 0.003775 \\
F4  &    2.9103399 $\pm$ 2.57e-05   & 0.000196  $\pm$  6.95e-06     &  0.909752  $\pm$ 0.005629 \\
F5  &    14.600872 $\pm$ 2.29e-05   & 0.000220  $\pm$  6.95e-06     &  0.302139  $\pm$ 0.005028 \\
F6  &    6.1912872 $\pm$ 3.80e-05   & 0.000132  $\pm$  6.95e-06     &  0.503004  $\pm$ 0.008320 \\
F7  &    17.927163 $\pm$ 3.40e-05   & 0.000148  $\pm$  6.95e-06     &  0.184552  $\pm$ 0.007456 \\
F8  &    3.6437002 $\pm$ 2.43e-05   & 0.000186  $\pm$  6.95e-06     &  0.365693  $\pm$ 0.005334 \\
F9  &    3.6416541 $\pm$ 2.63e-05   & 0.000171  $\pm$  6.95e-06     &  0.452262  $\pm$ 0.005751 \\
F10 &    4.3695361 $\pm$ 3.45e-05   & 0.000145  $\pm$  6.95e-06     &  0.722033  $\pm$ 0.007552 \\
F11 &    11.434968 $\pm$ 3.05e-05   & 0.000142  $\pm$  6.95e-06     &  0.250010  $\pm$ 0.006684 \\
F12 &    6.9172551 $\pm$ 3.66e-05   & 0.000134  $\pm$  6.95e-06     &  0.044063  $\pm$ 0.008006 \\
F13 &    5.8248381 $\pm$ 2.87e-05   & 0.000155  $\pm$  6.95e-06     &  0.795300  $\pm$ 0.006280 \\
F14 &    5.8186998 $\pm$ 2.93e-05   & 0.000149  $\pm$  6.95e-06     &  0.047370  $\pm$ 0.006408 \\
F15 &    15.666161 $\pm$ 3.99e-05   & 0.000126  $\pm$  6.95e-06     &  0.034572  $\pm$ 0.008745 \\
\hline
\end{tabular}

\label{tab:TIC8247_freqs}
\end{table*}

\begin{table*}[h!]
\centering
\small
\caption{Frequencies used to compute the Least-Squares Fourier fitting for the target TIC~126945917}

\begin{tabular}{lcccc}

Label & Frequency ($\mathrm{d^{-1}}$) & Amplitude & Phase \\
\hline
F1 & 	22.073819 $\pm$ 3.56e-06     & 0.000993  $\pm$  1.18e-05    &   0.107520 $\pm$   0.001889 \\
F2 & 	0.7297407 $\pm$ 3.34e-06     & 0.001045  $\pm$  1.18e-05    &   0.380749 $\pm$   0.001771 \\
F3 & 	0.7737380 $\pm$ 4.43e-06     & 0.000776  $\pm$  1.18e-05    &   0.909715 $\pm$   0.002353 \\
F4 & 	31.171173 $\pm$ 4.82e-06     & 0.000731  $\pm$  1.18e-05    &   0.326682 $\pm$   0.002557 \\
F5 & 	16.794981 $\pm$ 4.78e-06     & 0.000740  $\pm$  1.18e-05    &   0.472750 $\pm$   0.002536 \\
F6 & 	1.4907642 $\pm$ 4.02e-06     & 0.000864  $\pm$  1.18e-05    &   0.217327 $\pm$   0.002135 \\
F7 & 	14.714690 $\pm$ 4.87e-06     & 0.000726  $\pm$  1.18e-05    &   0.589102 $\pm$   0.002584 \\
F8 & 	30.541814 $\pm$ 4.56e-06     & 0.000771  $\pm$  1.18e-05    &   0.511148 $\pm$   0.002417 \\
F9 & 	26.217196 $\pm$ 6.25e-06     & 0.000567  $\pm$  1.18e-05    &   0.966106 $\pm$   0.003314 \\
F10& 	32.659786 $\pm$ 6.65e-06     & 0.000532  $\pm$  1.18e-05    &   0.491052 $\pm$   0.003526 \\
F11& 	6.7095688 $\pm$ 6.16e-06     & 0.000569  $\pm$  1.18e-05    &   0.957445 $\pm$   0.003267 \\
F12& 	30.545354 $\pm$ 7.98e-06     & 0.000438  $\pm$  1.18e-05    &   0.789437 $\pm$   0.004231 \\
F13& 	8.2021572 $\pm$ 7.77e-06     & 0.000449  $\pm$  1.18e-05    &   0.364410 $\pm$   0.004123 \\
F14& 	5.2192946 $\pm$ 7.81e-06     & 0.000447  $\pm$  1.18e-05    &   0.624241 $\pm$   0.004143 \\
F15& 	2.9529954 $\pm$ 8.31e-06     & 0.000428  $\pm$  1.18e-05    &   0.661682 $\pm$   0.004405 \\
F16& 	0.7011805 $\pm$ 8.17e-06     & 0.000415  $\pm$  1.18e-05    &   0.105064 $\pm$   0.004331 \\
F17& 	5.9735661 $\pm$ 7.23e-06     & 0.000486  $\pm$  1.18e-05    &   0.450736 $\pm$   0.003833 \\
F18& 	7.4621794 $\pm$ 8.77e-06     & 0.000400  $\pm$  1.18e-05    &   0.133543 $\pm$   0.004649 \\
F19& 	4.4964966 $\pm$ 8.12e-06     & 0.000424  $\pm$  1.18e-05    &   0.281985 $\pm$   0.004308 \\
F20& 	3.0124028 $\pm$ 8.57e-06     & 0.000405  $\pm$  1.18e-05    &   0.520629 $\pm$   0.004543 \\
F21& 	1.5297247 $\pm$ 1.05e-05     & 0.000319  $\pm$  1.18e-05    &   0.598491 $\pm$   0.005582 \\
F22& 	4.4486331 $\pm$ 1.01e-05     & 0.000347  $\pm$  1.18e-05    &   0.970609 $\pm$   0.005372 \\
F23& 	25.274629 $\pm$ 1.03e-05     & 0.000342  $\pm$  1.18e-05    &   0.767988 $\pm$   0.005467 \\
F24& 	13.294523 $\pm$ 1.02e-05     & 0.000346  $\pm$  1.18e-05    &   0.027543 $\pm$   0.005417 \\
F25& 	26.278536 $\pm$ 1.10e-05     & 0.000319  $\pm$  1.18e-05    &   0.703953 $\pm$   0.005850 \\
F26& 	9.6994829 $\pm$ 1.06e-05     & 0.000329  $\pm$  1.18e-05    &   0.239402 $\pm$   0.005659 \\
F27& 	0.8187155 $\pm$ 1.08e-05     & 0.000344  $\pm$  1.18e-05    &   0.300980 $\pm$   0.005729 \\
F28& 	1.4616867 $\pm$ 1.02e-05     & 0.000344  $\pm$  1.18e-05    &   0.577823 $\pm$   0.005417 \\
F29& 	3.7196275 $\pm$ 1.16e-05     & 0.000297  $\pm$  1.18e-05    &   0.636138 $\pm$   0.006194 \\
F30& 	27.277788 $\pm$ 1.24e-05     & 0.000284  $\pm$  1.18e-05    &   0.744833 $\pm$   0.006588 \\
\hline
\end{tabular}

\label{tab:TIC1269_freqs}
\end{table*}

\begin{table*}[h!]
\centering
\small
\caption{Frequencies used to compute the Least-Squares Fourier fitting for the target TIC~10756751}

\begin{tabular}{lcccc}

Label & Frequency ($\mathrm{d^{-1}}$) & Amplitude & Phase \\
\hline
F1  &   12.920924  $\pm$ 1.65e-06     & 0.001171  $\pm$ 6.49e-06     &    0.499714  $\pm$  0.0008 \\
F2  &   12.639542  $\pm$ 2.74e-06     & 0.000706  $\pm$ 6.49e-06     &    0.481234  $\pm$  0.0014 \\
F3  &   12.924281  $\pm$ 7.03e-06     & 0.000250  $\pm$ 6.49e-06     &    0.882769  $\pm$  0.0037 \\
F4  &   2.0107661  $\pm$ 8.07e-06     & 0.000221  $\pm$ 6.49e-06     &    0.668651  $\pm$  0.0043 \\
F5  &   15.074520  $\pm$ 1.11e-05     & 0.000172  $\pm$ 6.49e-06     &    0.450698  $\pm$  0.0059 \\
F6  &   2.8653349  $\pm$ 7.00e-06     & 0.000264  $\pm$ 6.49e-06     &    0.460549  $\pm$  0.0037 \\
F7  &   2.2893068  $\pm$ 7.81e-06     & 0.000232  $\pm$ 6.49e-06     &    0.857032  $\pm$  0.0041 \\
F8  &   1.7169598  $\pm$ 8.97e-06     & 0.000191  $\pm$ 6.49e-06     &    0.208902  $\pm$  0.0047 \\
F9  &   3.4386022  $\pm$ 1.00e-05     & 0.000194  $\pm$ 6.49e-06     &    0.886262  $\pm$  0.0053 \\
F10 &   2.5840063  $\pm$ 1.15e-05     & 0.000160  $\pm$ 6.49e-06     &    0.083443  $\pm$  0.0061 \\
F11 &   4.0173640  $\pm$ 1.35e-05     & 0.000146  $\pm$ 6.49e-06     &    0.642853  $\pm$  0.0072 \\
F12 &   12.408042  $\pm$ 1.53e-05     & 0.000122  $\pm$ 6.49e-06     &    0.534624  $\pm$  0.0081 \\
F13 &   1.7189086  $\pm$ 1.25e-05     & 0.000128  $\pm$ 6.49e-06     &    0.409329  $\pm$  0.0066 \\
F14 &   2.2939623  $\pm$ 1.44e-05     & 0.000113  $\pm$ 6.49e-06     &    0.364557  $\pm$  0.0077 \\
F15 &   2.0142847  $\pm$ 1.95e-05     & 0.000009  $\pm$ 6.49e-06     &    0.880057  $\pm$  0.0104 \\
F16 &   19.719587  $\pm$ 2.47e-05     & 0.000007  $\pm$ 6.49e-06     &    0.663119  $\pm$  0.0132 \\
\hline
\end{tabular}

\label{tab:TIC1075_freqs}
\end{table*}

\begin{table*}[h!]
\centering
\small
\caption{Frequencies used to compute the Least-Squares Fourier fitting for the target TIC~37817410}

\begin{tabular}{lcccc}

Label & Frequency ($\mathrm{d^{-1}}$) & Amplitude & Phase \\
\hline
F1  &   18.718979 $\pm$ 1.50e-06     & 0.003378  $\pm$  1.17e-05    &      0.880115	 $\pm$  0.000554 \\
F2  &   19.611191 $\pm$ 4.44e-06     & 0.001059  $\pm$  1.08e-05    &      0.934496	 $\pm$  0.001634 \\
F3  &   30.664819 $\pm$ 6.36e-06     & 0.000737  $\pm$  1.08e-05    &      0.146838  $\pm$  0.002340 \\
F4  &   1.5341865 $\pm$ 7.70e-06     & 0.000685  $\pm$  1.15e-05    &      0.600564  $\pm$  0.002855 \\
F5  &   1.1545303 $\pm$ 8.75e-06     & 0.000583  $\pm$  1.14e-05    &      0.029044  $\pm$  0.003227 \\
F6  &   25.429543 $\pm$ 1.17e-05     & 0.000397  $\pm$  1.08e-05    &      0.316141  $\pm$  0.004338 \\
F7  &   40.727537 $\pm$ 1.39e-05     & 0.000336  $\pm$  1.08e-05    &      0.334563  $\pm$  0.005116 \\
F8  &   24.520260 $\pm$ 1.45e-05     & 0.000320  $\pm$  1.08e-05    &      0.318793  $\pm$  0.005365 \\
F9  &   1.9097716 $\pm$ 1.41e-05     & 0.000358  $\pm$  1.14e-05    &      0.093028  $\pm$  0.005247 \\
F10 &   24.557621 $\pm$ 1.99e-05     & 0.000244  $\pm$  1.13e-05    &      0.974350  $\pm$  0.007436 \\
F11 &   20.451530 $\pm$ 1.74e-05     & 0.000268  $\pm$  1.08e-05    &      0.441621  $\pm$  0.006415 \\
F12 &   18.981561 $\pm$ 1.68e-05     & 0.000280  $\pm$  1.09e-05    &      0.799792  $\pm$  0.006189 \\
F13 &   0.3416380 $\pm$ 1.86e-05     & 0.000273  $\pm$  1.12e-05    &      0.597371  $\pm$  0.006818 \\
F14 &   2.3015752 $\pm$ 2.26e-05     & 0.000223  $\pm$  1.16e-05    &      0.080970  $\pm$  0.008362 \\
F15 &   23.791677 $\pm$ 2.34e-05     & 0.000207  $\pm$  1.13e-05    &      0.751740  $\pm$  0.008717 \\
F16 &   19.575011 $\pm$ 2.41e-05     & 0.000196  $\pm$  1.08e-05    &      0.266198	 $\pm$  0.008892 \\
F17 &   18.422385 $\pm$ 2.44e-05     & 0.000198  $\pm$  1.09e-05    &      0.329637	 $\pm$  0.008963 \\
F18 &   0.2565404 $\pm$ 2.44e-05     & 0.000191  $\pm$  1.10e-05    &      0.611227  $\pm$  0.008964 \\
F19 &   18.693240 $\pm$ 2.44e-05     & 0.000210  $\pm$  1.17e-05    &      0.199131  $\pm$  0.008976 \\
F20 &   0.4231242 $\pm$ 2.68e-05     & 0.000186  $\pm$  1.13e-05    &      0.163392  $\pm$  0.009849 \\
F21 &   45.158968 $\pm$ 2.88e-05     & 0.000162  $\pm$  1.08e-05    &      0.854718  $\pm$  0.010601 \\
\hline
\end{tabular}

\label{tab:TZEri_freqs}
\end{table*}

\end{document}